\documentclass[preprint,superscriptaddress,floatfix,sort&compress]{revtex4}

\usepackage[dvips]{graphicx}
\usepackage{amssymb,amsfonts,amsmath}
\usepackage{color}
\usepackage[font=scriptsize]{caption}
\usepackage[position=top,caption=false]{subfig}

\usepackage{ulem}

\newcommand{\1}{\begin{equation}}
\newcommand{\2}{\end{equation}}

\begin{document}

\title{Non-equilibrium chromosome looping via molecular slip-links}

\author{C. A. Brackley$^{1}$, J. Johnson$^{1}$, D. Michieletto$^{1}$, A.~N. Morozov$^1$, M. Nicodemi$^2$, P.~R. Cook$^3$, D. Marenduzzo$^1$}
\affiliation{$^1$ SUPA, School of Physics and Astronomy, University of Edinburgh, Peter Guthrie Tait Road, Edinburgh, EH9 3FD, UK, $^2$ Dipartimento di Fisica, Universita' di Napoli Federico II, INFN Napoli, CNR, SPIN, Complesso Universitario di Monte Sant'Angelo, 80126 Naples, Italy, $^3$ Sir William Dunn School of Pathology, University of Oxford, South Parks Road, Oxford, OX1 3RE, UK} 

\begin{abstract}
We propose a model for the formation of chromatin loops based on the diffusive sliding of a DNA-bound factor which can dimerise to form a molecular slip-link. Our slip-links mimic the behaviour of cohesin-like molecules, which, along with the CTCF protein, stabilize loops which organize the genome. By combining 3D Brownian dynamics simulations and 1D exactly solvable non-equilibrium models, we show that diffusive sliding is sufficient to account for the strong bias in favour of convergent CTCF-mediated chromosome loops observed experimentally. Importantly, our model does not require any underlying, and energetically costly, motor activity of cohesin. 
We also find that the diffusive motion of multiple slip-links along chromatin may be rectified by an intriguing ratchet effect that arises if slip-links bind to the chromatin at a preferred ``loading site''. This emergent collective behaviour is driven by a 1D osmotic pressure which is set up near the loading point, and favours the extrusion of loops which are much larger than the ones formed by single slip-links.
\end{abstract}

\maketitle

The formation of long-range contacts, or loops, within DNA and chromosomes is a process which critically affects gene expression~\cite{Alberts2014,Chambeyron2004}. For instance, looping between specific regulatory elements, such as  enhancers and promoters, can dramatically  increase transcription rates in eukaryotes~\cite{Alberts2014}. The formation of these loops can often be successfully predicted by equilibrium polymer physics models, which balance the energetic gain of protein-mediated interactions with the entropic loss associated with loop formation~\cite{Marenduzzo2006,Hanke2003,Vilar2006}. 

\begin{figure}[t!]
	\centering
	\includegraphics[width=0.8\textwidth]{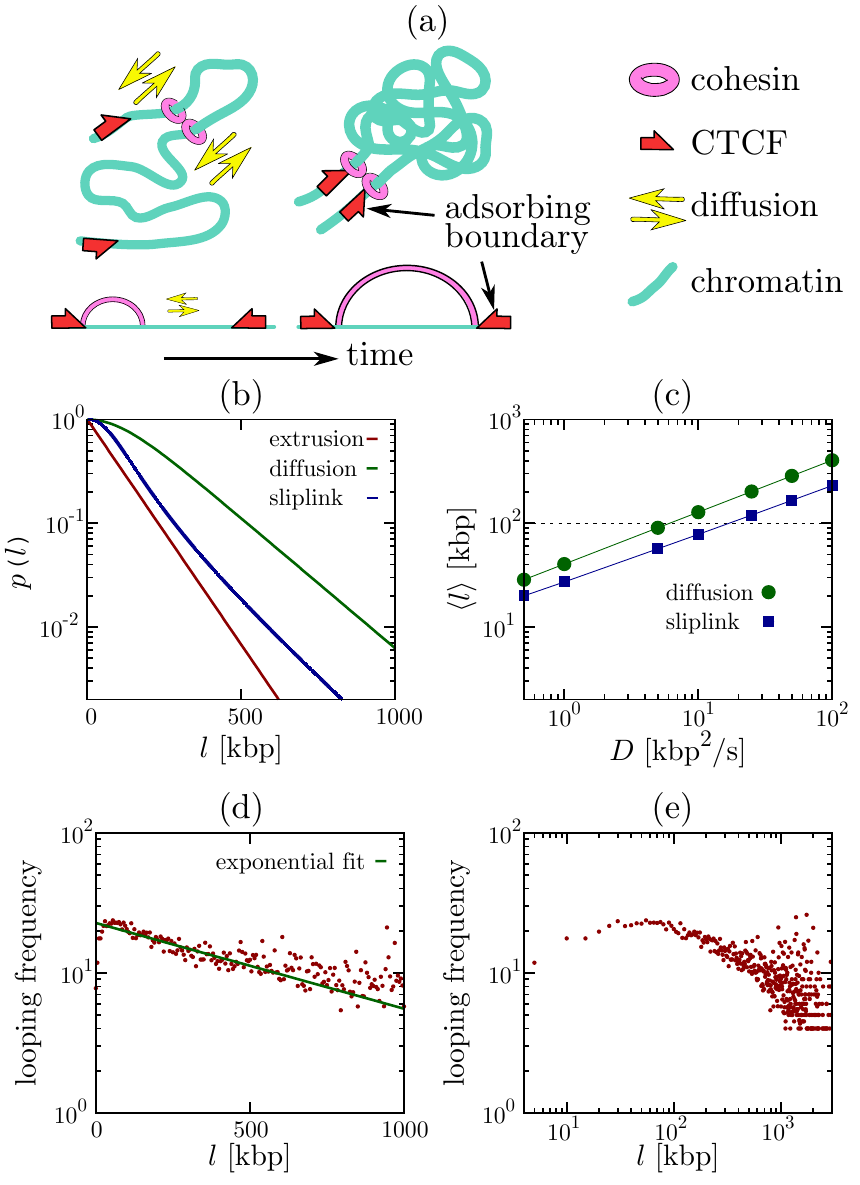}
	\caption{Nonequilibrium chromosome looping and CTCF-mediated loops. (a) Schematic of our model of cohesins as diffusing slip-links (see text for details). (b) Probability of nonequilibrium loop formation in exactly solvable 1D models as a function of loop size $l$. Curves correspond to models involving (i) extrusion, (ii) diffusion and (iii) slip-links. Parameters are $k_{\rm off}^{-1}=20$~min and: (i) $v=5$ kbp/min; (ii, iii) $D_0=0.01$ $\mu$m$^2$/s$^{-1}$ and a compaction of $50$ bp/nm (so $D=D_0C^2=2.5$ kbp$^2$/s); (iii) $\sigma_{\rm sl}=1$ kbp, and $c=1$~\cite{Mirny2011}. In all cases, $p(l)$ decays exponentially, contrary to the power law decay which is characteristic of polymer loop formation in equilibrium. (c) Average loop size for models involving diffusion and slip-links. Parameters are as in (b), apart from $D$ which is varied. Both models can account for the formation of a typical CTCF loop of $100$ kbp (dotted line) provided $D\sim 10$ kbp$^2$/s or above. (d,e) Analysis of ChIA-PET data for contacts within a Mbp, where both the adjacent segments in a contact are bound to CTCF~\cite{Tang2015}. The decay of the looping probability is better fitted as exponential (d), rather than a power law (e).}
	\label{fig1}
\end{figure}

However, recent high-throughput chromosome conformation capture (``Hi-C'') experiments~\cite{Dekker2002,Rao2014} have fundamentally challenged the view that equilibrium physics is sufficient to model chromosome looping. Hi-C experiments showed that the genomes of most eukaryotic organisms are partitioned into domains -- called ``topologically associated domains'', or TADs. In several cases, these domains were found to be enclosed within a chromosome loop, $100-1000$ kilo-basepairs (kpb) in size, and the bases of the loops are statistically enriched in binding sites for the CCCTC-binding factor (CTCF)~\cite{Rao2014,Dixon2012}. CTCF is a DNA-binding protein with an important role in gene regulation, 
and CTCF-mediated loops preferentially enclose inducible genes, which are normally silent and are pressed into action in response to a stimulus (e.g., an inflammation or an increased concentration of a morphogen during development)~\cite{Oti2016}. The DNA-binding motif of CTCF is not palindromic, meaning that it has a specific direction on the DNA. Surprisingly, Hi-C analyses have recently revealed that most of the CTCF binding sequences only form a loop when they are in a ``convergent'' orientation (Fig.~\ref{fig1}a)~\cite{Rao2014,DeWit2015}. Very few contacting CTCFs have a ``parallel'' orientation, and virtually none have a ``divergent'' one. This strong bias is puzzling, because, if we imagine drawing arrows on the chromatin fiber (corresponding to the CTCF binding site directions), then two loops with a pair of convergent or divergent arrows at their base are compatible with the same 3D structure~\cite{Rao2014,Oti2016}. Consequently, no equilibrium polymer physics model can possibly distinguish between the two patterns.  

In most cases CTCF-mediated loops are associated with cohesin~\cite{Uhlmann2016}, a ring-like protein complex thought to bind DNA by topologically embracing it~\cite{Veld2014}. There are two popular models for how cohesin might achieve this -- as a dimer acting as molecular ``hand-cuffs'' in which each ring embraces one DNA duplex (Fig.~\ref{fig1}a), or as a single ring that embraces two duplexes~\cite{Nasmyth2011}. In both cases, the dimer/ring acts as a sliding bridge or molecular slip-link~\cite{Metzler2002,Michieletto2016softmatter}, and we will use the latter term to describe both cases. \textit{In vitro} and \textit{in vivo} experiments show that cohesin does indeed topologically link to DNA (with binding  mediated by ``loader proteins'' such as Scc2 or NIPBL~\cite{Alberts2014,Stigler2016}), that it can slide along DNA diffusively, and that it remains bound for $\tau\sim 20$ minutes before dissociating~\cite{Ocampo-Hafalla2011,Stigler2016,Gerlich2006,Ladurner2014,Hansen2016}.

One recent attempt to address the mechanism underlying CTCF and cohesin-mediated looping is the ``loop extrusion model'' which argues that cohesins (or other ``loop extruding factors'') can actively create loops of $100-1000$ kbp by travelling in opposite directions along the chromosome~\cite{Alipour2012,Fudenberg2016,Sanborn2015}. This model is appealing because it naturally explains the bias in favour of convergent loops, if the slip-link gets stuck when it finds a CTCF binding site pointing towards it (an assumption consistent with experiments probing CTCF and cohesin binding~\cite{Oti2016,Fudenberg2016,topoIIcohesinCTCF}). However, the model is based on several assumptions for which experimental evidence is currently lacking: most notably it requires that (i) each cohesin is able to determine and maintain the correct direction in order to extrude (rather than shrink) a loop, and (ii) that cohesin must be able to extrude loops of $100-1000$ kbp in a timescale $\sim\tau$. The extrusion speed would therefore need to exceed that of an RNA polymerase (which is $v\sim 1$ kbp/min), one of the most efficient and processive known chromosome-bound motors active during interphase. Whilst cohesin is known to have ATPase activity, this seems not to be involved in directional motion; instead it drives the gate-opening mechanism needed to form a topologically stable association with DNA~\cite{Nasmyth2011}.

Here, we propose an alternative model for the formation of CTCF-mediated loops, which does not require unidirectional motion, or any energetically costly explicit bias favouring loop extrusion. We start from the observation that the molecular topology of cohesin dimers -- i.e. that of a slip-link -- is compatible with diffusive sliding along DNA or chromatin~\cite{Ocampo-Hafalla2011}. From this premise, we formulate a non-equilibrium model where the binding and unbinding kinetics of cohesin violates detailed balance, and show that within this context passive sliding is sufficient to account for both the creation of loops of hundreds of kbp before dissociation, and the formation of convergent CTCF-mediated loops. The probability of formation of such loops in our framework differs from the canonical power law decay governing the statistics of equilibrium polymer looping, and is consistent with currently available data on CTCF loops. Finally, we show that many-body steric interactions between diffusing slip-links which always bind close to a preferred ``loading site'' can lead to the emergence of an ``osmotic ratchet'' which promotes loop extrusion over shrinking, again in the absence of any bias in the microscopic molecular diffusion. 

\section*{Results}

\subsection{Single slip-link, 1D model} 

We begin by discussing an exactly solvable 1D model where a slip-link consisting of two cohesin rings in a dimer slides along the chromatin fiber. We assume that the slip-link binds with the cohesin rings at adjacent positions on the fiber (as in~\cite{Fudenberg2016}), and that there is a constant detachment rate $k_{\rm off} \sim \tau^{-1}$. We consider two CTCF proteins bound to the fiber at a separation $l$ to create a convergent pair of CTCF binding sites. [The case of a divergent pair is treated in the SI, and as expected leads to no stable looping (Fig.~S1).] As cohesin interacts with a CTCF in a directionality-dependent manner (only when it faces the CTCF binding motif~\cite{Oti2016,Fudenberg2016}) we assume that when the slip-link reaches the two convergent CTCF sites it undergoes a conformational change decreasing $k_{\rm off}$~\footnote{Note $k_{\rm off}$ will decrease, whilst remaining finite, even with a purely thermodynamic (directional) attraction between cohesin and CTCF. The crucial non-equilibrium aspect, to yield a non-power-law looping probability is that, when (re)binding, the monomers are {\it always} adjacent along the fiber.}. For simplicity, we allow the rings forming one cohesin to diffuse until their separation reaches $l$, or until the dimer spontaneously unbinds, and consider both to be absorbing states. This is a non-equilibrium model as the binding-unbinding kinetics violate detailed balance: this violation is consistent with the experimentally well-established~\cite{Nasmyth2011} ATPase activity associated with cohesin-chromatin interactions.

At a given time $t$, the slip-link holds together a chromatin loop of size $x(t)$. In order to take into account the entropic loss associated with this loop, we include an effective thermodynamic potential $V(x)$ (detailed below). The probability that the cohesin holds a loop of size $x$ at time $t$, obeys the following generalised Fokker-Plank equation, 
\begin{equation}\label{FPE}
\frac{\partial p(x,t)}{\partial t}=-k_{\rm off}p(x,t)+\frac{\partial}{\partial x}\left[\frac{1}{\gamma}\frac{dV}{dx}p(x,t)\right]+D\frac{\partial^2}{\partial x^2}p(x,t),
\end{equation}
where $D$ and $\gamma$ are the effective diffusion and drag coefficients describing the relative motion between chromatin and cohesins. The fluctuation-dissipation theorem implies $D=k_BT/\gamma$. The initial condition for Eq.~(\ref{FPE}) is $p(x,0)=\delta(x-\sigma_{\rm sl})$, where $\sigma_{\rm sl}$ is the size of the slip-link. 
 Boundary conditions are reflecting at $x=\sigma_{\rm sl}$ and absorbing at $x=l$. 

We consider three possible cases. First, we model the ``loop extrusion'' process proposed in~\cite{Alipour2012,Goloborodko2016,Fudenberg2016} by setting $D=0$ and $\frac{1}{\gamma}\frac{dV}{dx}=v$, where $v$ is the extrusion speed. Second, we consider a ``diffusion'' model where cohesin diffuses in the absence of a potential, $V=0$. Third, we model the effect of chromatin looping on a diffusing cohesin dimer by setting $V(x)=ck_BT\log{(x)}$, which models the thermodynamic entropic cost of looping via the known contact (looping) probability $p_{\rm eq}(x)\sim x^{-c}$. In this formula, $c$ is a universal exponent, which in 3D is equal to $1.5$ for loops made by infinitesimally thin random walks~\cite{Duplantier1989}, $\sim 2.1$ for internal looping within self-avoiding chains~\cite{Duplantier1989,Carlon2002}, and $1$ for contacts within a ``fractal globule''~\cite{Mirny2011}. We refer to this third model, with a logarithmic looping potential, as the ``slip-link'' model, as it more closely resembles the dynamics of slip-links on polymers~\cite{Metzler2002,Hanke2003,Michieletto2016softmatter}.

As detailed in the SI, we can analytically find the probability that a cohesin dimer binding at $t=0$ will, at some point, form a CTCF-mediated loop before detaching. Denoting this probability by $p(l)$, the three models predict the following dependence on loop size $l$ (Fig. 1b),
\begin{eqnarray}\label{loopingprobabilities}
p_{\rm extr}(l) & = & e^{-k_{\rm off}l/v} \\ \nonumber
p_{\rm diff}(l) & = & \frac{1}{{\cosh(\alpha l)}} \\ \nonumber
p_{\rm slip}(l) & = & \left(\frac{l}{\sigma_{\rm sl}}\right)^n
\frac{I_{m-1}(\alpha l)K_m(\alpha l)+I_m(\alpha l)K_{m-1}(\alpha l)}{I_{m-1}(\alpha \sigma_{\rm sl})K_m(\alpha l)+I_m(\alpha l)K_{m-1}(\alpha\sigma_{\rm sl})}
\end{eqnarray}
where $\alpha=\sqrt{k_{\rm off}/D}$, $n=(1-c)/2$, and $m=(1+c)/2$; $I$ and $K$ denote the modified Bessel functions of the first and second kind respectively. Note that we have taken the $\sigma_{\rm sl}\to 0$ limit for the loop extrusion ($p_{\rm extr}(l)$) and diffusion ($p_{\rm diff}(l)$) cases. 

For large $l$, Eqs.~(\ref{loopingprobabilities}) predict exponential decay of CTCF-mediated looping probabilities for all cases (Fig. 1b), with a power law correction for slip-links, $p_{\rm slip}(l) \sim e^{-\alpha l}l^{-c/2}$. This is markedly different from the power laws which determine the looping probability of an equilibrium polymer~\cite{Metzler2002,Michieletto2016softmatter}.
The decay length is $v/k_{\rm off}$ for the loop extrusion model~\cite{Fudenberg2016}, and $\alpha^{-1}=\sqrt{D/k_{\rm off}}$ for the diffusion and slip-link models; these are therefore the typical looping lengths formed before cohesin detaches. A typical CTCF-mediated loop length \textit{in vivo} is $\sim100$kbp~\cite{Rao2014,Oti2016}; taking $\tau=20$~min means that loop extrusion is viable if $v>5$~kbp/min or more (compare $v=1$~kbp/min for polymerase), whereas the diffusion or slip-link models require $D\sim 10$~kbp$^2$/s or above. The latter condition is achievable under normal conditions: for instance, assuming a diffusion coefficient $D_0=0.01-0.1$~$\mu$m$^2$s$^{-1}$, reasonable for protein sliding on chromatin~\cite{facilitateddiffusion}, and a compaction rate $C=50$~bp/nm (intermediate between a $10$ and a $30$ nm chromatin fiber) we get $D=D_0C^2\sim 25-250$~kbp$^2$/s (Figs. 1b, 1c).

Hi-C experiments measuring the frequency of contacts between all genomic loci can be used to infer chromatin looping probabilities, and largely support a power law decay of contacts, with a chromosome-dependent exponent whose average is $~1$~\cite{Barbieri2012}. However, these data do not distinguish between CTCF-mediated loops and other contacts, which may either form stochastically or through other chromatin-binding proteins~\cite{Barbieri2012,Brackley2013a,Brackley2016nar,Chiarriello2016}. Chromatin Interaction Analysis by Paired-End Tag Sequencing (ChIA-PET) experiments~\cite{Oti2016} are able to single out contacts where both anchor points are bound to a protein of interest. Intriguingly, in CTCF ChIA-PET data~\cite{Tang2015}, a fit to an exponential leads to a reasonable decay length (typical loop size) equal to $\sim 500-1000$ kbp (Fig.~1d, and SI, Fig.~S1). On the contrary, a fit to a power law is poorer and yields an effective exponent which is far from those which may be expected from equilibrium polymer physics models (Fig.~1e, and SI, Figs.~S8, S9).  
{This simple analysis supports the idea that the statistics of CTCF-mediated loops retain a signature of their underlying non-equilibrium nature and that, remarkably, this feature is captured by our simple 1D model of diffusing slip-links.}

\subsection{Single slip-link, 3D simulations} 

We now ask whether the effects predicted by our simple 1D model are confirmed by 3D Brownian dynamics simulations, which can more accurately account for both the 3D structure of chromosomal loops and the steric interactions between a molecular slip-link and chromatin. Specifically, we enquire whether diffusive sliding may account for the formation of CTCF-mediated loops, and what the probability of formation of such loops is in a 3D simulation. We consider a chromatin fiber modelled as a bead-and-spring polymer with bead diameter $\sigma=30$ nm, $C=100$ bp/nm, and vary the persistence length $l_p$ (see SI for further details and results); we load a single slip-link, modelled by two rigid rings (each of the ring has diameter $2R\sim 3.4 \sigma$, and thickness $\sigma_{\rm sl}=\sigma$) linked via a semiflexible hinge, which favours a planar hand-cuff configuration with the centre of the rings a distance $2R$ apart (Fig.~2a, inset, and SI, Fig.~S5). 

\begin{figure}[t!]
	\centering
        \includegraphics[width=0.8\textwidth]{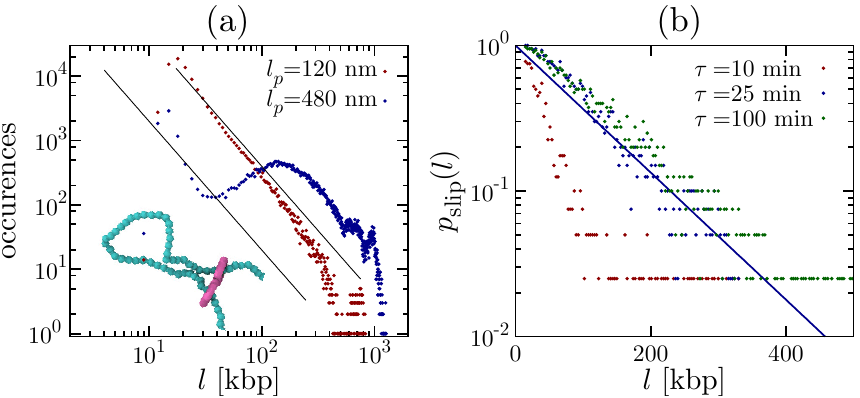}
	\caption{Brownian dynamics simulations of a molecular slip-link. (a) Frequency with which the slip-link is associated with a loop of size $l$ (equilibrium model with $k_{\rm off}=0$). Black lines show a power law with exponent $-2$. Inset: snapshot of a configuration of a slip-link diffusing along a chromatin fiber with persistence length $12\sigma$.
(b) Non-equilibrium looping probability for a slip-link, computed from Brownian dynamics trajectories, with different $k_{\rm off}^{-1}$. The blue line shows an exponential fit to the $k_{\rm off}^{-1}=25$ min data.}
	\label{fig2}
\end{figure}

Figure~2a shows the frequency with which the slip-links form loops of size $l$ when $k_{\rm off}=0$ and the persistence length of the fiber takes values equal to either $4\sigma$ or $16\sigma$ (see SI for more values of $l_p$). Since the slip-link cannot unbind, these curves represent equilibrium looping probabilities, $p_{\rm eq}(l)$; they indeed show clear evidence of power law decay for large $l$ (Fig.~2a, and SI, Fig.~S6). The exponent is $c\sim 2$, in line with the contact probability between internal points in a self-avoiding walk in a good solvent~\cite{Carlon2002,Duplantier1989,Hanke2003} (appropriate for a segment of open chromatin). 
Each of the curves in Fig.~2a also shows two peaks at small and intermediate $l$. The first peak is associated with the minimum loop length needed to bring two beads $\sim 2R$ apart. The second one is due to the competition between the energy required to bend the chromatin fiber and entropy of loop formation. This behaviour is recapitulated by an analysing the distribution function of internal distances of a semi-flexible polymer (see SI, Fig.~S6).

An important question is to what extent this more accurate 3D description can account for CTCF-mediated looping spanning several hundreds of kbp under realistic values of $k_{\rm off}$. In Figure~2b we plot the probability  $p_{\rm slip}(l)$ that a slip-link reaches a separation $l$ along a flexible chromatin fiber. 
In other words, we ask, as in our 1D model in Figure~\ref{fig1}, whether diffusing cohesin dimers can reach CTCF sites separated by a distance $l$ before disassociating. Our simulations predict that such loops can indeed form; for instance a $100$ kbp loop can form with probability $\sim 0.3$ with $k_{\rm off}^{-1}=\tau=25$~min (see also SI, Suppl. Movie 1 and Fig.~S7). 
We highlight that, in agreement with our 1D non-equilibrium models, the decay of $p_{\rm slip}(l)$ is only compatible with an exponential, rather than a power law decay typically found in equilibrium simulations (Fig.~2a). 

\subsection{Multiple slip-links and the osmotic ratchet} 

So far, we have considered the case of a single cohesin dimer diffusing on chromatin. When, instead, multiple slip-links are present on the same chromatin segment, they may interact either sterically or entropically.

We first modify our 1D model to simulate the stochastic dynamics of $N$ slip-links diffusing along a chromatin fiber of size $L$ (Figs. 3a-c, see Methods), discretized into segments of length $\sigma_{\rm sl}$. Each slip-link can exist in an unbound or chromatin-bound state while the binding and unbinding rates are $k_{\rm on}$ and $k_{\rm off}$ respectively. When binding, the two slip-link monomers always occupy neighbouring sites along the fiber. [For simplicity, we set $k_{\rm on}=k_{\rm off}$; then, the number of bound and unbound slip-links is equal in steady state.] 
There are excluded volume interactions between cohesins, such that the ends of the slip-links cannot cross each other. In the SI, we present further results which include a ``looping weight'' (Figs.~S3 and S4), i.e. an effective potential which accounts for the entropy, or probability of formation, of a network of loops~\cite{Metzler2002,Michieletto2016softmatter}. This effective potential has a quantitative effect on our results, but it does not modify the qualitative trends; in this section we report findings from 1D stochastic simulations without looping weight, as this simpler version is simpler to analyse theoretically.

\begin{figure}[t!]
	\centering
	\includegraphics[width=0.8\textwidth]{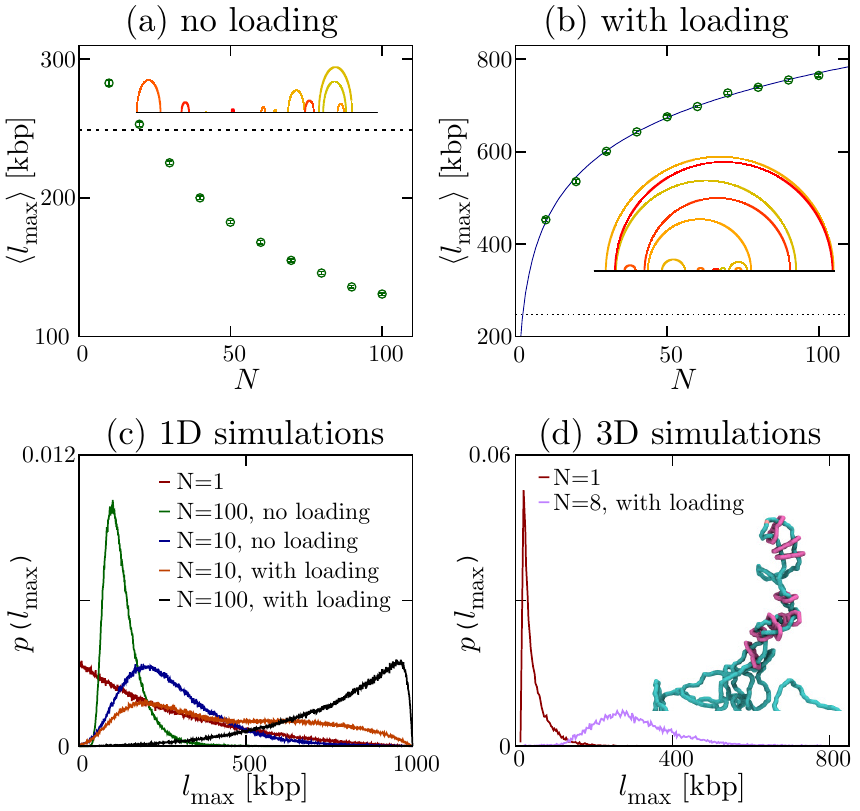}
	\caption{When multiple slip-links bind at a loading site, they set up an osmotic ratchet to yield effective loop extrusion. (a,b) Results from 1D simulations of diffusing slip-links which rebind either randomly (a), or always at a loading site (b): plots show the time average of the largest loop for the case without ``looping weight'' (the case with looping weight is shown in the SI, Fig.~S4). Parameters are: $\sigma_{\rm sl}=1$ kbp, $L=1000$ kbp, $k_{\rm on}^{-1}=k_{\rm off}^{-1}=25$ min, the diffusion coefficient  of a monomer is $D\sim 33.35$ kbp$^2$/s, while the number of slip-links, $N$ is varied. There are reflecting boundary conditions at the two ends of the fiber. Typical configurations for $N=20$ are shown as insets. The dotted line in (a,b) denotes the average loop size with a single slip-link (which is placed back on the fiber every time it detaches); the solid line in (b) is a fit to $a+b \log{N}$ (see text). (c) Distribution probability of the size of the largest loop for different values of $N$, and different models (with/without loading, see legend; the case with loading and with looping weight is shown in Fig.~S4). (d) Results from 3D Brownian dynamics simulations of multiple slip-links binding at a loading site, for a chromatin fiber of $L=3000$ kbp, with $k_{\rm off}^{-1}=25$ min. The plots show the distribution probability of the  size of the largest loop for $N=1$ (with $k_{\rm on}\to \infty$), and $N=8$ (with $k_{\rm on}=10 k_{\rm off}$). A typical snapshot showing stacking of nested loops and effective extrusion for $N=8$ is shown as an inset.}
	\label{fig3}
\end{figure}

We consider two cases: in the first one slip-links bind at random (unoccupied) locations on the fiber; in the second one binding occurs at a preferred ``loading site''. Figure~\ref{fig3}a shows the time average of the maximal loop size $\langle l_{\rm max} \rangle$ in steady state as a function of $N$, for the case of random rebinding. As the fiber gets more crowded, the slip-links form consecutive loops (Fig.~\ref{fig3}a, inset, and Fig.~S3c) which compete with each other. The maximum loop size which can be formed is thus limited and we observe that $\langle l_{\rm max} \rangle$ decreases steadily with $N$ (Fig.~\ref{fig3}a). 

A strikingly different result is found when slip-links always bind to the same location. This scenario mimics the experimental finding that the topological association of cohesin to DNA is facilitated by a loader protein (e.g., Scc2 or NIPBL~\cite{Alberts2014,Stigler2016}), which has preferential binding sites within the genome. 
In this case, we observe that the maximum loop size $\langle l_{\rm max} \rangle$ {\it increases} with $N$, rather than decreasing (Fig.~\ref{fig3}b), thereby favouring effective loop growth over shrinking. In other words, the system now works as a ratchet, which rectifies the diffusion of the two ends of the loop subtended by a slip-link. The typical loop network found in steady state is very different from the case of random rebinding, and now consists of a large proportion of nested loops (Fig.~\ref{fig3}b, inset, and Fig.~S3d), which reinforce each other rather than competing for space along the fiber. Figure~\ref{fig3}c shows the probability distribution of sizes for the largest loop and confirms the dramatic difference between the cases with and without preferential ``loading site''. 

In order to fully address this intriguing ratchet effect, we performed 3D Brownian dynamics simulations of a chromatin fiber interacting with $N$ slip-links which can bind and unbind, in the presence of a loading site (see Methods). We find that the cooperative behaviour of multiple slip-links loaded at a specific site again leads to ratcheting and, in particular, we find that the outer loops can easily span hundreds of kbp (Fig.~\ref{fig3}d) even when considering only a few slip-links. This ratchet effect may therefore provide a microscopic basis for the loop extrusion model in~\cite{Alipour2012,Fudenberg2016,Sanborn2015}, valid under conditions where several cohesins (or other molecular slip-links) are bound to the same chromatin region. 

The inset in Figure~\ref{fig3}d shows a typical snapshot of our 3D simulations, which also highlights that nested loops are formed by closely ``stacked'' slip-links (Fig.~\ref{fig3}d, inset, and Fig.~S3e) that can be easily recognised in arc-diagram representations by characteristic ``rainbow'' patterns (SI, Suppl. Movie 2). Stacking is triggered by entropic forces which tend to diminish the total number of loops~\cite{Metzler2002} and, thus, cluster the slip-links together. This behaviour is reminiscent of the ``bridging-induced attraction''~\cite{Brackley2013a,Brackley2016nar} which drives the formation of protein clusters, although the underlying mechanism is here purely entropic.
 
To understand the emergence of a self-organized ratchet, we construct a simple theory by further analysing the 1D model (without looping weight). The key factor is the existence of a non-uniform slip-link density, and hence an osmotic pressure; the associated pressure gradient creates a force that rectifies the motion of cohesin rings placed close to the loading site. If volume exclusion does not significantly affect the density or pressure profiles (an assumption which is true if $N\sigma_{\rm sl}/L$ is small enough, and which holds in our 1D stochastic simulations, Fig.~S2), we can write down the following phenomenological equation determining the size of a loop subtended by a symmetrically progressing slip-link starting from the loader
\begin{equation}\label{ratchet}
\frac{dl}{dt}=-2 k_BT \sigma_{\rm sl} \left[\frac{\partial \rho}{\partial x}\right]_{x=l/2}
= k_{\rm on} N_{\rm off}\sigma_{\rm sl}e^{-\alpha l/2},
\end{equation}
where $N_{\rm off}=N k_{\rm off}/(k_{\rm on}+k_{\rm off})$ is the average number of unbound cohesins. 

The maximal speed of this ``osmotic ratchet'' is achieved for loops close to the loading site, and is $v\sim k_{\rm on}N_{\rm off}\sigma_{\rm sl}$, which holds for $k_{\rm on}\sigma_{\rm sl}^2/D\ll 1$. The maximal possible ratchet speed, achieved for $k_{\rm on}\sigma_{\rm sl}^2/D\gg 1$, is instead $v\sim D/\sigma_{\rm sl}$, similar to the case of a ``Brownian ratchet'' modelling actin polymerisation close to a fluctuating membrane~\cite{Peskin1993}. Eq.~(\ref{ratchet}) further predicts that at a given time, $l$ should grow logarithmically with $N$, and our data are indeed fitted well by the functional form $a+b\log{N}$ (Fig.~\ref{fig3}b). While the theory we have presented explains why the case with and without loading are fundamentally different, and why the former can create a ratchet, we should not expect it to be quantitative as it describes the dynamics of a typical loop, rather than the largest one, which is considered in Figure~\ref{fig3}. In this respect, a more refined theory would require the application of extreme value statistics to a problem of $N$ random walkers~\cite{Majumdar2010}: it would be of interest to pursue this analysis in the future. 

\begin{figure}[t!]
	\centering
	\includegraphics[width=0.8\textwidth]{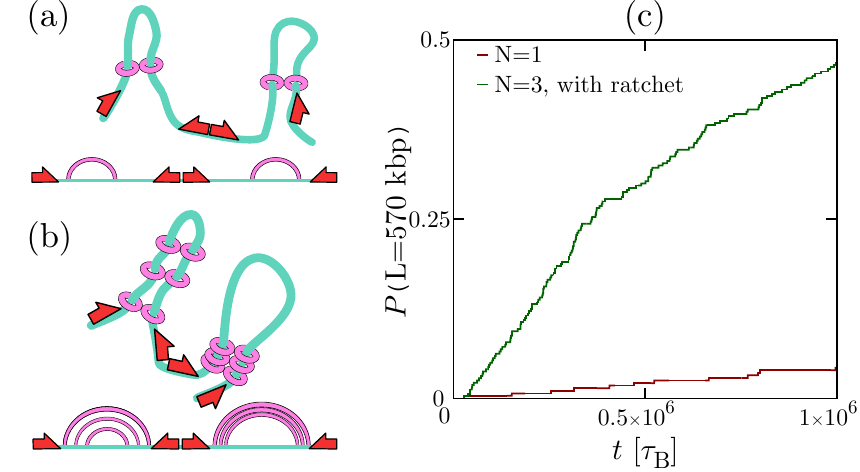}
\caption{The osmotic ratchet can substantially increase the probability of CTCF-mediated loop formation. (a) A chromatin fiber of size $L=2000\sigma$ is divided up into sections of $190\sigma$ (corresponding to $570$ kbp). Each section is limited by a convergent pair of CTCF sites, which strongly binds the slip-link. A single (permanently bound) slip-link is placed in each section. (b) Same as (a), but now there are three slip-links in each section, forming three nested loops. (c) This plot shows the probability that the largest loop has reached the convergent CTCF sites, as a function of time (measured here in Brownian times, each corresponding to $\sim 0.1$s). Curves for different $l_p$ are shown in the SI (Fig.~S7).}
\label{fig4}
\end{figure}

\section*{Discussion and conclusions} 

In summary, we have proposed a series of non-equilibrium models to study the dynamics of molecular slip-links which bind to and detach from a chromatin fiber, and can slide diffusively in 1D along it when bound. These slip-links are a model for cohesins, condensins or other structurally similar proteins which bind DNA by topologically embracing it. 

We suggest that these slip-links may play a pivotal role in the dynamic organization of chromosome loops. First, we have shown that 1D diffusive sliding of cohesin~\cite{Stigler2016} is sufficient to explain the experimentally observed bias favouring the looping of convergent CTCF binding sites over divergent ones. The only additional assumptions are that the binding kinetics violate detailed balance, and that cohesin and CTCF interact in a directional manner, in agreement with experimental evidence~\cite{Oti2016,Fudenberg2016,topoIIcohesinCTCF}. 
Second, we have found that the probability of formation of cohesin/CTCF-mediated loop follows an exponential decay; hence it is fundamentally different from the power laws which govern polymer looping in thermodynamic equilibrium.
Third, we have shown that a non-trivial and self-organized collective behaviour emerges when multiple slip-links slide along the same chromatin region. In particular, when binding occurs preferentially at a ``loading site'', a ratchet effect arises, where slip-links set up an osmotic pressure which rectifies molecular diffusion. This ratchet provides a viable microscopic mechanism yielding extrusion of chromatin loops, which has been postulated by recent models of chromatin organization~\cite{Fudenberg2016,Sanborn2015}.

Each of these results depends critically on our assumption that the binding/unbinding interactions between slip-links and chromatin violate detailed balance so that the system is out of equilibrium. This assumption is consistent with the current understanding of cohesin, which displays an ATPase activity associated with conformational changes.
Importantly, our results show that the formation of convergent CTCF-mediated loops spanning several hundreds of kbp is consistent with simple unbiased diffusion at the molecular level, in the absence of any background motor activity or unidirectional motion of the slip-links, as was required in the previous loop extrusion model~\cite{Fudenberg2016}. 

A consequence of our work is that it poses well-defined constraints on the minimal cohesin diffusion coefficient, $D_0$ (in $\mu$m$^2$/s) and chromatin compaction, $C$ (in bp/nm), which are needed for slip-links to be able to organise chromosome loops of hundreds of kbp, such as the typical convergent CTCF-mediated loops found in mammalian genomes~\cite{Rao2014}. Specifically, our analysis shows that a single slip-link requires a value of $D_0 C^2\sim 10$ kbp$^2$/s or more, to reach the end of a $100$ kbp loop. The worst possible case for our theory occurs if the substrate along which cohesin slides is decompacted, as covering the same genomic stretch will then require larger $D_0$. This scenario, corresponding to $C\sim 10$ bp/nm (i.e., a $10$-nm fiber), requires $D_0\sim 0.1$ $\mu$m$^2$/s for loops to form in practice. This value is achievable for 1D protein diffusion along chromatin~\cite{facilitateddiffusion}. 

Recent \textit{in vitro} experiments of cohesin diffusion on naked DNA~\cite{Stigler2016} have been used to extrapolate a slow diffusion rate of  $D_0\sim 0.001$ $\mu$m$^2$/s on chromatin. If this is the case \textit{in vivo},  a single slip-link can still create CTCF-mediated loops, but only if cohesin slides on a compact 30-nm-like fiber ($C\sim 100$ bp/nm). In practice, though, the ratchet effect described in Fig.~\ref{fig3} would enhance diffusivity to facilitate loop  formation (e.g. $\sim10$ cohesin molecules can effectively increase $D_0$ by an order of magnitude). A second factor which may favour the formation of longer loops is that during the S and G2 phase of the cell cycle there is evidence of a subpopulation of cohesins with $\tau \gg 20$ min. To quantify these arguments, Fig.~\ref{fig4} shows the probability of formation of a long $570$ kbp convergent CTCF-mediated loop over time, for chromatin fibers with different $N$ (see also Fig.~S7). The results show that the osmotic ratchet is at work with as few as $3$ bound cohesins per loop, and dramatically enhances looping probability (Fig.~\ref{fig4}c). 

We hope that the findings reported in this work will prompt new studies to measure cohesin diffusion accurately on reconstituted chromatin fibers \textit{in vitro}, and as a function of the number of cohesins bound to the fiber. This would allow a test of our osmotic ratchet in the lab, and at the same time determine whether our mechanism of loop formation based on unbiased diffusive sliding can work {\it in vivo}. Furthermore, we envisage that high-throughput (e.g., ChIA-PET) experiments probing the looping probabilities controlled by different loop-mediating proteins (such as PolII) will also help illuminate general non-equilibrium features of loop formation in chromosomes. Finally, from a theoretical point of view, it would be of interest to study the behaviour of chromatin fibers subject to both molecular slip-links and more conventional bridging and writing proteins, such as those considered in~\cite{Barbieri2012,Brackley2013a,Brackley2016nar,Chiarriello2016, Michieletto2016PRX}.

\section*{Methods}

\subsection*{3D Brownian dynamics simulations of molecular slip-links} 
In our Brownian dynamics simulations we follow the evolution of a chromatin fibre and of one or more slip-links which are topologically bound to the fibre. The dynamics are evolved using the LAMMPS software~\cite{lammps} in Brownian dynamics mode (see SI). 

Briefly, the force field includes: (i) non-linear springs between neighbouring chromatin beads to ensure chain connectivity; (ii) bending rigidity of the chromatin fiber; (iii) excluded volume interactions between any two beads, representing part of either the chromatin fiber or any slip-link. The main novelty in our simulations is represented by the slip-links, which are modeled as a pair of rings, each of which moves as a rigid body. The two rings are kept together by non-linear springs, and there are bending interactions favouring the ``open'' hand-cuff configuration (see SI, Fig.~S5). 

Slip-links can detach (when on the chromatin) or bind (when detached), at rates $k_{\rm off}$ and $k_{\rm on}$ respectively. Stochastic detachment/binding are simulating by means of an external code, which is interfaced with LAMMPS and is called every $1000$ Brownian dynamics steps. More details are given in the SI.

\subsection*{1D stochastic simulations of molecular slip-links}
1D stochastic simulations of $N$ diffusing slip-link dimers on a chromatin fiber were performed by using a kinetic Monte-Carlo algorithm, where rules were defined as follows. At each time step, on average, we attempt to randomly move, either to the left or to the right, each of the monomers in slip-links which are bound to chromatin. Moves which would lead to clash between any two monomers are rejected. In the case with looping weight (see SI), we also include a Metropolis acceptance test, with an effective potential which mimics the entropic weight associated with the instantaneous looping network. At each time step, we also attempt to rebind on average each detached slip-links, and detach each bound slip-links, with rates $k_{\rm on}$ and $k_{\rm off}$ respectively; rebinding occurs either at a random position or at a loading site in the middle of the chromatin fiber. 

\section*{Acknowledgements}

This work was supported by ERC (CoG 648050, THREEDCELLPHYSICS), by ISCRA Grants HP10CYFPS5 and HP10CRTY8P, by computer resources at INFN and Scope at the University of Naples, and by the Einstein BIH Fellowship Award to MN.

\bibliographystyle{apsrev}
\bibliography{rings}

\begin{thebibliography}{39}
\expandafter\ifx\csname natexlab\endcsname\relax\def\natexlab#1{#1}\fi
\expandafter\ifx\csname bibnamefont\endcsname\relax
  \def\bibnamefont#1{#1}\fi
\expandafter\ifx\csname bibfnamefont\endcsname\relax
  \def\bibfnamefont#1{#1}\fi
\expandafter\ifx\csname citenamefont\endcsname\relax
  \def\citenamefont#1{#1}\fi
\expandafter\ifx\csname url\endcsname\relax
  \def\url#1{\texttt{#1}}\fi
\expandafter\ifx\csname urlprefix\endcsname\relax\def\urlprefix{URL }\fi
\providecommand{\bibinfo}[2]{#2}
\providecommand{\eprint}[2][]{\url{#2}}

\bibitem[{\citenamefont{Alberts et~al.}(2014)\citenamefont{Alberts, Johnson,
  Lewis, Morgan, and Raff}}]{Alberts2014}
\bibinfo{author}{\bibfnamefont{B.}~\bibnamefont{Alberts}},
  \bibinfo{author}{\bibfnamefont{A.}~\bibnamefont{Johnson}},
  \bibinfo{author}{\bibfnamefont{J.}~\bibnamefont{Lewis}},
  \bibinfo{author}{\bibfnamefont{D.}~\bibnamefont{Morgan}}, \bibnamefont{and}
  \bibinfo{author}{\bibfnamefont{M.}~\bibnamefont{Raff}},
  \emph{\bibinfo{title}{{Molecular Biology of the Cell}}}
  (\bibinfo{publisher}{Taylor {\&} Francis}, \bibinfo{year}{2014}), ISBN
  \bibinfo{isbn}{0815344643}.

\bibitem[{\citenamefont{Chambeyron and Bickmore}(2004)}]{Chambeyron2004}
\bibinfo{author}{\bibfnamefont{S.}~\bibnamefont{Chambeyron}} \bibnamefont{and}
  \bibinfo{author}{\bibfnamefont{W.~A.} \bibnamefont{Bickmore}},
  \bibinfo{journal}{Curr. Opin. Cell Biol.} \textbf{\bibinfo{volume}{16}},
  \bibinfo{pages}{256} (\bibinfo{year}{2004}).

\bibitem[{\citenamefont{Marenduzzo et~al.}(2006)\citenamefont{Marenduzzo,
  Micheletti, and Cook}}]{Marenduzzo2006}
\bibinfo{author}{\bibfnamefont{D.}~\bibnamefont{Marenduzzo}},
  \bibinfo{author}{\bibfnamefont{C.}~\bibnamefont{Micheletti}},
  \bibnamefont{and} \bibinfo{author}{\bibfnamefont{P.~R.} \bibnamefont{Cook}},
  \bibinfo{journal}{Biophys. J.} \textbf{\bibinfo{volume}{90}},
  \bibinfo{pages}{3712} (\bibinfo{year}{2006}).

\bibitem[{\citenamefont{Hanke and Metzler}(2003)}]{Hanke2003}
\bibinfo{author}{\bibfnamefont{A.}~\bibnamefont{Hanke}} \bibnamefont{and}
  \bibinfo{author}{\bibfnamefont{R.}~\bibnamefont{Metzler}},
  \bibinfo{journal}{Biophys. J.} \textbf{\bibinfo{volume}{85}},
  \bibinfo{pages}{167} (\bibinfo{year}{2003}).

\bibitem[{\citenamefont{Vilar and Saiz}(2006)}]{Vilar2006}
\bibinfo{author}{\bibfnamefont{J.~M.~G.} \bibnamefont{Vilar}} \bibnamefont{and}
  \bibinfo{author}{\bibfnamefont{L.}~\bibnamefont{Saiz}},
  \bibinfo{journal}{Phys. Rev. Lett.} \textbf{\bibinfo{volume}{96}},
  \bibinfo{pages}{238103} (\bibinfo{year}{2006}).

\bibitem[{\citenamefont{Dekker et~al.}(2002)\citenamefont{Dekker, Rippe,
  Dekker, and Kleckner}}]{Dekker2002}
\bibinfo{author}{\bibfnamefont{J.}~\bibnamefont{Dekker}},
  \bibinfo{author}{\bibfnamefont{K.}~\bibnamefont{Rippe}},
  \bibinfo{author}{\bibfnamefont{M.}~\bibnamefont{Dekker}}, \bibnamefont{and}
  \bibinfo{author}{\bibfnamefont{N.}~\bibnamefont{Kleckner}},
  \bibinfo{journal}{Science} \textbf{\bibinfo{volume}{295}},
  \bibinfo{pages}{1306} (\bibinfo{year}{2002}).

\bibitem[{\citenamefont{Rao et~al.}(2014)\citenamefont{Rao, Huntley, Durand,
  Stamenova, Bochkov, Robinson, Sanborn, Machol, Omer, Lander
  et~al.}}]{Rao2014}
\bibinfo{author}{\bibfnamefont{S.~S.~P.} \bibnamefont{Rao}},
  \bibinfo{author}{\bibfnamefont{M.~H.} \bibnamefont{Huntley}},
  \bibinfo{author}{\bibfnamefont{N.~C.} \bibnamefont{Durand}},
  \bibinfo{author}{\bibfnamefont{E.~K.} \bibnamefont{Stamenova}},
  \bibinfo{author}{\bibfnamefont{I.~D.} \bibnamefont{Bochkov}},
  \bibinfo{author}{\bibfnamefont{J.~T.} \bibnamefont{Robinson}},
  \bibinfo{author}{\bibfnamefont{A.~L.} \bibnamefont{Sanborn}},
  \bibinfo{author}{\bibfnamefont{I.}~\bibnamefont{Machol}},
  \bibinfo{author}{\bibfnamefont{A.~D.} \bibnamefont{Omer}},
  \bibinfo{author}{\bibfnamefont{E.~S.} \bibnamefont{Lander}},
  \bibnamefont{et~al.}, \bibinfo{journal}{Cell} \textbf{\bibinfo{volume}{159}},
  \bibinfo{pages}{1665} (\bibinfo{year}{2014}), \eprint{1206.5533}.

\bibitem[{\citenamefont{Dixon et~al.}(2012)\citenamefont{Dixon, Selvaraj, Yue,
  Kim, Li, Shen, Hu, Liu, and Ren}}]{Dixon2012}
\bibinfo{author}{\bibfnamefont{J.~R.} \bibnamefont{Dixon}},
  \bibinfo{author}{\bibfnamefont{S.}~\bibnamefont{Selvaraj}},
  \bibinfo{author}{\bibfnamefont{F.}~\bibnamefont{Yue}},
  \bibinfo{author}{\bibfnamefont{A.}~\bibnamefont{Kim}},
  \bibinfo{author}{\bibfnamefont{Y.}~\bibnamefont{Li}},
  \bibinfo{author}{\bibfnamefont{Y.}~\bibnamefont{Shen}},
  \bibinfo{author}{\bibfnamefont{M.}~\bibnamefont{Hu}},
  \bibinfo{author}{\bibfnamefont{J.~S.} \bibnamefont{Liu}}, \bibnamefont{and}
  \bibinfo{author}{\bibfnamefont{B.}~\bibnamefont{Ren}},
  \bibinfo{journal}{Nature} \textbf{\bibinfo{volume}{485}},
  \bibinfo{pages}{376} (\bibinfo{year}{2012}).

\bibitem[{\citenamefont{Oti et~al.}(2016)\citenamefont{Oti, Falck, Huynen, and
  Zhou}}]{Oti2016}
\bibinfo{author}{\bibfnamefont{M.}~\bibnamefont{Oti}},
  \bibinfo{author}{\bibfnamefont{J.}~\bibnamefont{Falck}},
  \bibinfo{author}{\bibfnamefont{M.~A.} \bibnamefont{Huynen}},
  \bibnamefont{and} \bibinfo{author}{\bibfnamefont{H.}~\bibnamefont{Zhou}},
  \bibinfo{journal}{BMC genomics} \textbf{\bibinfo{volume}{17}},
  \bibinfo{pages}{252} (\bibinfo{year}{2016}).

\bibitem[{\citenamefont{de~Wit et~al.}(2015)\citenamefont{de~Wit, Vos,
  Holwerda, Valdes-Quezada, Verstegen, Teunissen, Splinter, Wijchers, Krijger,
  and de~Laat}}]{DeWit2015}
\bibinfo{author}{\bibfnamefont{E.}~\bibnamefont{de~Wit}},
  \bibinfo{author}{\bibfnamefont{E.~S.~M.} \bibnamefont{Vos}},
  \bibinfo{author}{\bibfnamefont{S.~J.~B.} \bibnamefont{Holwerda}},
  \bibinfo{author}{\bibfnamefont{C.}~\bibnamefont{Valdes-Quezada}},
  \bibinfo{author}{\bibfnamefont{M.~J. A.~M.} \bibnamefont{Verstegen}},
  \bibinfo{author}{\bibfnamefont{H.}~\bibnamefont{Teunissen}},
  \bibinfo{author}{\bibfnamefont{E.}~\bibnamefont{Splinter}},
  \bibinfo{author}{\bibfnamefont{P.~J.} \bibnamefont{Wijchers}},
  \bibinfo{author}{\bibfnamefont{P.~H.~L.} \bibnamefont{Krijger}},
  \bibnamefont{and} \bibinfo{author}{\bibfnamefont{W.}~\bibnamefont{de~Laat}},
  \bibinfo{journal}{Molecular Cell} \textbf{\bibinfo{volume}{60}},
  \bibinfo{pages}{676} (\bibinfo{year}{2015}).

\bibitem[{\citenamefont{Uhlmann}(2016)}]{Uhlmann2016}
\bibinfo{author}{\bibfnamefont{F.}~\bibnamefont{Uhlmann}},
  \bibinfo{journal}{Nat. Rev. Mol. Cell. Biol.} \textbf{\bibinfo{volume}{17}}
  (\bibinfo{year}{2016}).

\bibitem[{\citenamefont{Huis in~'t Veld et~al.}(2014)\citenamefont{Huis in~'t
  Veld, Herzog, Ladurner, Davidson, Piric, Kreidl, Bhaskara, Aebersold, and
  Peters}}]{Veld2014}
\bibinfo{author}{\bibfnamefont{P.~J.} \bibnamefont{Huis in~'t Veld}},
  \bibinfo{author}{\bibfnamefont{F.}~\bibnamefont{Herzog}},
  \bibinfo{author}{\bibfnamefont{R.}~\bibnamefont{Ladurner}},
  \bibinfo{author}{\bibfnamefont{I.~F.} \bibnamefont{Davidson}},
  \bibinfo{author}{\bibfnamefont{S.}~\bibnamefont{Piric}},
  \bibinfo{author}{\bibfnamefont{E.}~\bibnamefont{Kreidl}},
  \bibinfo{author}{\bibfnamefont{V.}~\bibnamefont{Bhaskara}},
  \bibinfo{author}{\bibfnamefont{R.}~\bibnamefont{Aebersold}},
  \bibnamefont{and} \bibinfo{author}{\bibfnamefont{J.~M.}
  \bibnamefont{Peters}}, \bibinfo{journal}{Science}
  \textbf{\bibinfo{volume}{346}}, \bibinfo{pages}{968} (\bibinfo{year}{2014}).

\bibitem[{\citenamefont{Nasmyth}(2011)}]{Nasmyth2011}
\bibinfo{author}{\bibfnamefont{K.}~\bibnamefont{Nasmyth}},
  \bibinfo{journal}{Nat. Cell Biol.} \textbf{\bibinfo{volume}{13}},
  \bibinfo{pages}{1170} (\bibinfo{year}{2011}).

\bibitem[{\citenamefont{Metzler et~al.}(2002)\citenamefont{Metzler, Hanke,
  Dommersnes, Kantor, and Kardar}}]{Metzler2002}
\bibinfo{author}{\bibfnamefont{R.}~\bibnamefont{Metzler}},
  \bibinfo{author}{\bibfnamefont{A.}~\bibnamefont{Hanke}},
  \bibinfo{author}{\bibfnamefont{P.~G.} \bibnamefont{Dommersnes}},
  \bibinfo{author}{\bibfnamefont{Y.}~\bibnamefont{Kantor}}, \bibnamefont{and}
  \bibinfo{author}{\bibfnamefont{M.}~\bibnamefont{Kardar}},
  \bibinfo{journal}{Phys. Rev. E} \textbf{\bibinfo{volume}{65}},
  \bibinfo{pages}{1} (\bibinfo{year}{2002}).

\bibitem[{\citenamefont{Michieletto}(2016)}]{Michieletto2016softmatter}
\bibinfo{author}{\bibfnamefont{D.}~\bibnamefont{Michieletto}},
  \bibinfo{journal}{Soft Matter} \textbf{\bibinfo{volume}{12}},
  \bibinfo{pages}{9485} (\bibinfo{year}{2016}).

\bibitem[{\citenamefont{Stigler et~al.}(2016)\citenamefont{Stigler,
  {\c{C}}amdere, Koshland, and Greene}}]{Stigler2016}
\bibinfo{author}{\bibfnamefont{J.}~\bibnamefont{Stigler}},
  \bibinfo{author}{\bibfnamefont{G.}~\bibnamefont{{\c{C}}amdere}},
  \bibinfo{author}{\bibfnamefont{D.~E.} \bibnamefont{Koshland}},
  \bibnamefont{and} \bibinfo{author}{\bibfnamefont{E.~C.}
  \bibnamefont{Greene}}, \bibinfo{journal}{Cell Reports}
  \textbf{\bibinfo{volume}{15}}, \bibinfo{pages}{988} (\bibinfo{year}{2016}).

\bibitem[{\citenamefont{Ocampo-Hafalla and Uhlmann}(2011)}]{Ocampo-Hafalla2011}
\bibinfo{author}{\bibfnamefont{M.~T.} \bibnamefont{Ocampo-Hafalla}}
  \bibnamefont{and} \bibinfo{author}{\bibfnamefont{F.}~\bibnamefont{Uhlmann}},
  \bibinfo{journal}{J. Cell. Sci.} \textbf{\bibinfo{volume}{124}},
  \bibinfo{pages}{685} (\bibinfo{year}{2011}).

\bibitem[{\citenamefont{Gerlich et~al.}(2006)\citenamefont{Gerlich, Koch,
  Dupeux, Peters, and Ellenberg}}]{Gerlich2006}
\bibinfo{author}{\bibfnamefont{D.}~\bibnamefont{Gerlich}},
  \bibinfo{author}{\bibfnamefont{B.}~\bibnamefont{Koch}},
  \bibinfo{author}{\bibfnamefont{F.}~\bibnamefont{Dupeux}},
  \bibinfo{author}{\bibfnamefont{J.~M.} \bibnamefont{Peters}},
  \bibnamefont{and}
  \bibinfo{author}{\bibfnamefont{J.}~\bibnamefont{Ellenberg}},
  \bibinfo{journal}{Curr. Biol.} \textbf{\bibinfo{volume}{16}},
  \bibinfo{pages}{1571} (\bibinfo{year}{2006}).

\bibitem[{\citenamefont{Ladurner et~al.}(2014)\citenamefont{Ladurner, Bhaskara,
  in't Veld, Davidson, Kreidl, Petzold, and Peters}}]{Ladurner2014}
\bibinfo{author}{\bibfnamefont{R.}~\bibnamefont{Ladurner}},
  \bibinfo{author}{\bibfnamefont{V.}~\bibnamefont{Bhaskara}},
  \bibinfo{author}{\bibfnamefont{P.~J.~H.} \bibnamefont{in't Veld}},
  \bibinfo{author}{\bibfnamefont{I.~F.} \bibnamefont{Davidson}},
  \bibinfo{author}{\bibfnamefont{E.}~\bibnamefont{Kreidl}},
  \bibinfo{author}{\bibfnamefont{G.}~\bibnamefont{Petzold}}, \bibnamefont{and}
  \bibinfo{author}{\bibfnamefont{J.~M.} \bibnamefont{Peters}},
  \bibinfo{journal}{Curr. Biol.} \textbf{\bibinfo{volume}{24}},
  \bibinfo{pages}{2228} (\bibinfo{year}{2014}).

\bibitem[{\citenamefont{Hansen et~al.}(2016)\citenamefont{Hansen, Pustova,
  Cattoglio, Tjian, and Darzacq}}]{Hansen2016}
\bibinfo{author}{\bibfnamefont{A.~S.} \bibnamefont{Hansen}},
  \bibinfo{author}{\bibfnamefont{I.}~\bibnamefont{Pustova}},
  \bibinfo{author}{\bibfnamefont{C.}~\bibnamefont{Cattoglio}},
  \bibinfo{author}{\bibfnamefont{R.}~\bibnamefont{Tjian}}, \bibnamefont{and}
  \bibinfo{author}{\bibfnamefont{X.}~\bibnamefont{Darzacq}},
  \bibinfo{journal}{bioRxiv}  (\bibinfo{year}{2016}),
  \eprint{http://www.biorxiv.org/content/early/2016/12/12/093476.full.pdf}.

\bibitem[{\citenamefont{Alipour and Marko}(2012)}]{Alipour2012}
\bibinfo{author}{\bibfnamefont{E.}~\bibnamefont{Alipour}} \bibnamefont{and}
  \bibinfo{author}{\bibfnamefont{J.~F.} \bibnamefont{Marko}},
  \bibinfo{journal}{Nucleic Acids Res.} \textbf{\bibinfo{volume}{40}},
  \bibinfo{pages}{11202} (\bibinfo{year}{2012}).

\bibitem[{\citenamefont{Fudenberg et~al.}(2016)\citenamefont{Fudenberg,
  Imakaev, Lu, Goloborodko, Abdennur, and Mirny}}]{Fudenberg2016}
\bibinfo{author}{\bibfnamefont{G.}~\bibnamefont{Fudenberg}},
  \bibinfo{author}{\bibfnamefont{M.}~\bibnamefont{Imakaev}},
  \bibinfo{author}{\bibfnamefont{C.}~\bibnamefont{Lu}},
  \bibinfo{author}{\bibfnamefont{A.}~\bibnamefont{Goloborodko}},
  \bibinfo{author}{\bibfnamefont{N.}~\bibnamefont{Abdennur}}, \bibnamefont{and}
  \bibinfo{author}{\bibfnamefont{L.~A.} \bibnamefont{Mirny}},
  \bibinfo{journal}{Cell Reports} \textbf{\bibinfo{volume}{15}},
  \bibinfo{pages}{2038} (\bibinfo{year}{2016}).

\bibitem[{\citenamefont{Sanborn et~al.}(2015)\citenamefont{Sanborn, Rao, Huang,
  Durand, Huntley, Jewett, Bochkov, Chinnappan, Cutkosky, Li
  et~al.}}]{Sanborn2015}
\bibinfo{author}{\bibfnamefont{A.~L.} \bibnamefont{Sanborn}},
  \bibinfo{author}{\bibfnamefont{S.~S.~P.} \bibnamefont{Rao}},
  \bibinfo{author}{\bibfnamefont{S.-C.} \bibnamefont{Huang}},
  \bibinfo{author}{\bibfnamefont{N.~C.} \bibnamefont{Durand}},
  \bibinfo{author}{\bibfnamefont{M.~H.} \bibnamefont{Huntley}},
  \bibinfo{author}{\bibfnamefont{A.~I.} \bibnamefont{Jewett}},
  \bibinfo{author}{\bibfnamefont{I.~D.} \bibnamefont{Bochkov}},
  \bibinfo{author}{\bibfnamefont{D.}~\bibnamefont{Chinnappan}},
  \bibinfo{author}{\bibfnamefont{A.}~\bibnamefont{Cutkosky}},
  \bibinfo{author}{\bibfnamefont{J.}~\bibnamefont{Li}}, \bibnamefont{et~al.},
  \bibinfo{journal}{Proc. Natl. Acad. Sci. USA}  (\bibinfo{year}{2015}).

\bibitem[{\citenamefont{Uuskula-Reimand
  et~al.}(2016)\citenamefont{Uuskula-Reimand, Hou, Samavarchi-Tehrani,
  Vietri~Rudan, Liang, Medina-Rivera, Mohammed, Schmidt, Schwalie, Young
  et~al.}}]{topoIIcohesinCTCF}
\bibinfo{author}{\bibfnamefont{L.}~\bibnamefont{Uuskula-Reimand}},
  \bibinfo{author}{\bibfnamefont{H.}~\bibnamefont{Hou}},
  \bibinfo{author}{\bibfnamefont{P.}~\bibnamefont{Samavarchi-Tehrani}},
  \bibinfo{author}{\bibfnamefont{M.}~\bibnamefont{Vietri~Rudan}},
  \bibinfo{author}{\bibfnamefont{M.}~\bibnamefont{Liang}},
  \bibinfo{author}{\bibfnamefont{A.}~\bibnamefont{Medina-Rivera}},
  \bibinfo{author}{\bibfnamefont{H.}~\bibnamefont{Mohammed}},
  \bibinfo{author}{\bibfnamefont{D.}~\bibnamefont{Schmidt}},
  \bibinfo{author}{\bibfnamefont{P.}~\bibnamefont{Schwalie}},
  \bibinfo{author}{\bibfnamefont{E.~J.} \bibnamefont{Young}},
  \bibnamefont{et~al.}, \bibinfo{journal}{Gen. Biol.}
  \textbf{\bibinfo{volume}{17}}, \bibinfo{pages}{182} (\bibinfo{year}{2016}).

\bibitem[{\citenamefont{Mirny}(2011)}]{Mirny2011}
\bibinfo{author}{\bibfnamefont{L.~A.} \bibnamefont{Mirny}},
  \bibinfo{journal}{Chromosome Res.} \textbf{\bibinfo{volume}{19}},
  \bibinfo{pages}{37} (\bibinfo{year}{2011}).

\bibitem[{\citenamefont{Tang and coworkers}(2015)}]{Tang2015}
\bibinfo{author}{\bibfnamefont{Z.}~\bibnamefont{Tang}} \bibnamefont{and}
  \bibinfo{author}{\bibnamefont{coworkers}}, \bibinfo{journal}{Cell}
  \textbf{\bibinfo{volume}{163}}, \bibinfo{pages}{1611} (\bibinfo{year}{2015}).

\bibitem[{\citenamefont{Goloborodko et~al.}(2016)\citenamefont{Goloborodko,
  Imakaev, Marko, and Mirny}}]{Goloborodko2016}
\bibinfo{author}{\bibfnamefont{A.}~\bibnamefont{Goloborodko}},
  \bibinfo{author}{\bibfnamefont{M.~V.} \bibnamefont{Imakaev}},
  \bibinfo{author}{\bibfnamefont{J.~F.} \bibnamefont{Marko}}, \bibnamefont{and}
  \bibinfo{author}{\bibfnamefont{L.}~\bibnamefont{Mirny}},
  \bibinfo{journal}{eLife} pp. \bibinfo{pages}{1--20} (\bibinfo{year}{2016}).

\bibitem[{\citenamefont{Duplantier}(1989)}]{Duplantier1989}
\bibinfo{author}{\bibfnamefont{B.}~\bibnamefont{Duplantier}},
  \bibinfo{journal}{J. Stat. Phys.} \textbf{\bibinfo{volume}{54}},
  \bibinfo{pages}{581} (\bibinfo{year}{1989}).

\bibitem[{\citenamefont{Carlon et~al.}(2002)\citenamefont{Carlon, Orlandini,
  and Stella}}]{Carlon2002}
\bibinfo{author}{\bibfnamefont{E.}~\bibnamefont{Carlon}},
  \bibinfo{author}{\bibfnamefont{E.}~\bibnamefont{Orlandini}},
  \bibnamefont{and} \bibinfo{author}{\bibfnamefont{A.~L.}
  \bibnamefont{Stella}}, \bibinfo{journal}{Phys. Rev. Lett.}
  \textbf{\bibinfo{volume}{88}}, \bibinfo{pages}{198101}
  (\bibinfo{year}{2002}).

\bibitem[{\citenamefont{Normanno et~al.}(2015)\citenamefont{Normanno,
  Lydia~Boudarene, Dugast-Darzacq, Chen, Richter, Proux, Benichou, Voituriez,
  Darzacq, and Dahan}}]{facilitateddiffusion}
\bibinfo{author}{\bibfnamefont{D.}~\bibnamefont{Normanno}},
  \bibinfo{author}{\bibfnamefont{L.}~\bibnamefont{Lydia~Boudarene}},
  \bibinfo{author}{\bibfnamefont{C.}~\bibnamefont{Dugast-Darzacq}},
  \bibinfo{author}{\bibfnamefont{J.}~\bibnamefont{Chen}},
  \bibinfo{author}{\bibfnamefont{C.}~\bibnamefont{Richter}},
  \bibinfo{author}{\bibfnamefont{F.}~\bibnamefont{Proux}},
  \bibinfo{author}{\bibfnamefont{O.}~\bibnamefont{Benichou}},
  \bibinfo{author}{\bibfnamefont{R.}~\bibnamefont{Voituriez}},
  \bibinfo{author}{\bibfnamefont{X.}~\bibnamefont{Darzacq}}, \bibnamefont{and}
  \bibinfo{author}{\bibfnamefont{M.}~\bibnamefont{Dahan}},
  \bibinfo{journal}{Nat. Comm.} \textbf{\bibinfo{volume}{6}},
  \bibinfo{pages}{7357} (\bibinfo{year}{2015}).

\bibitem[{\citenamefont{Barbieri et~al.}(2012)\citenamefont{Barbieri, Chotalia,
  Fraser, Lavitas, Dostie, Pombo, and Nicodemi}}]{Barbieri2012}
\bibinfo{author}{\bibfnamefont{M.}~\bibnamefont{Barbieri}},
  \bibinfo{author}{\bibfnamefont{M.}~\bibnamefont{Chotalia}},
  \bibinfo{author}{\bibfnamefont{J.}~\bibnamefont{Fraser}},
  \bibinfo{author}{\bibfnamefont{L.-M.} \bibnamefont{Lavitas}},
  \bibinfo{author}{\bibfnamefont{J.}~\bibnamefont{Dostie}},
  \bibinfo{author}{\bibfnamefont{A.}~\bibnamefont{Pombo}}, \bibnamefont{and}
  \bibinfo{author}{\bibfnamefont{M.}~\bibnamefont{Nicodemi}},
  \bibinfo{journal}{Proc. Natl. Acad. Sci. USA} \textbf{\bibinfo{volume}{109}},
  \bibinfo{pages}{16173} (\bibinfo{year}{2012}).

\bibitem[{\citenamefont{Brackley et~al.}(2013)\citenamefont{Brackley, Taylor,
  Papantonis, Cook, and Marenduzzo}}]{Brackley2013a}
\bibinfo{author}{\bibfnamefont{C.~A.} \bibnamefont{Brackley}},
  \bibinfo{author}{\bibfnamefont{S.}~\bibnamefont{Taylor}},
  \bibinfo{author}{\bibfnamefont{A.}~\bibnamefont{Papantonis}},
  \bibinfo{author}{\bibfnamefont{P.~R.} \bibnamefont{Cook}}, \bibnamefont{and}
  \bibinfo{author}{\bibfnamefont{D.}~\bibnamefont{Marenduzzo}},
  \bibinfo{journal}{Proc. Natl. Acad. Sci. USA} \textbf{\bibinfo{volume}{110}},
  \bibinfo{pages}{E3605} (\bibinfo{year}{2013}).

\bibitem[{\citenamefont{Brackley et~al.}(2016)\citenamefont{Brackley, Johnson,
  Kelly, Cook, and Marenduzzo}}]{Brackley2016nar}
\bibinfo{author}{\bibfnamefont{C.~A.} \bibnamefont{Brackley}},
  \bibinfo{author}{\bibfnamefont{J.}~\bibnamefont{Johnson}},
  \bibinfo{author}{\bibfnamefont{S.}~\bibnamefont{Kelly}},
  \bibinfo{author}{\bibfnamefont{P.~R.} \bibnamefont{Cook}}, \bibnamefont{and}
  \bibinfo{author}{\bibfnamefont{D.}~\bibnamefont{Marenduzzo}},
  \bibinfo{journal}{Nucleic Acids Res.} \textbf{\bibinfo{volume}{44}},
  \bibinfo{pages}{3503} (\bibinfo{year}{2016}), \eprint{1511.01848}.

\bibitem[{\citenamefont{Chiariello et~al.}(2016)\citenamefont{Chiariello,
  Bianco, Annunziatella, Esposito, and Nicodemi}}]{Chiarriello2016}
\bibinfo{author}{\bibfnamefont{M.}~\bibnamefont{Chiariello}},
  \bibinfo{author}{\bibfnamefont{S.}~\bibnamefont{Bianco}},
  \bibinfo{author}{\bibfnamefont{C.}~\bibnamefont{Annunziatella}},
  \bibinfo{author}{\bibfnamefont{A.}~\bibnamefont{Esposito}}, \bibnamefont{and}
  \bibinfo{author}{\bibfnamefont{M.}~\bibnamefont{Nicodemi}},
  \bibinfo{journal}{Scientific Reports} \textbf{\bibinfo{volume}{6}},
  \bibinfo{pages}{29775} (\bibinfo{year}{2016}).

\bibitem[{\citenamefont{Peskin et~al.}(1993)\citenamefont{Peskin, Odell, and
  Oster}}]{Peskin1993}
\bibinfo{author}{\bibfnamefont{C.~S.} \bibnamefont{Peskin}},
  \bibinfo{author}{\bibfnamefont{G.~M.} \bibnamefont{Odell}}, \bibnamefont{and}
  \bibinfo{author}{\bibfnamefont{G.~F.} \bibnamefont{Oster}},
  \bibinfo{journal}{Biophys. J.} \textbf{\bibinfo{volume}{65}},
  \bibinfo{pages}{316} (\bibinfo{year}{1993}).

\bibitem[{\citenamefont{Majumdar et~al.}(2010)\citenamefont{Majumdar, Comtet,
  and Randon-F@urling}}]{Majumdar2010}
\bibinfo{author}{\bibfnamefont{S.}~\bibnamefont{Majumdar}},
  \bibinfo{author}{\bibfnamefont{A.}~\bibnamefont{Comtet}}, \bibnamefont{and}
  \bibinfo{author}{\bibfnamefont{J.}~\bibnamefont{Randon-F@urling}},
  \bibinfo{journal}{J. Stat. Phys.} \textbf{\bibinfo{volume}{138}},
  \bibinfo{pages}{955} (\bibinfo{year}{2010}).

\bibitem[{\citenamefont{Michieletto et~al.}(2016)\citenamefont{Michieletto,
  Orlandini, and Marenduzzo}}]{Michieletto2016PRX}
\bibinfo{author}{\bibfnamefont{D.}~\bibnamefont{Michieletto}},
  \bibinfo{author}{\bibfnamefont{E.}~\bibnamefont{Orlandini}},
  \bibnamefont{and}
  \bibinfo{author}{\bibfnamefont{D.}~\bibnamefont{Marenduzzo}},
  \bibinfo{journal}{Phys. Rev. X} \textbf{\bibinfo{volume}{6}},
  \bibinfo{pages}{041047} (\bibinfo{year}{2016}).

\bibitem[{\citenamefont{Plimpton}(1995)}]{lammps}
\bibinfo{author}{\bibfnamefont{S.}~\bibnamefont{Plimpton}},
  \bibinfo{journal}{J. Comput. Phys.} \textbf{\bibinfo{volume}{117}},
  \bibinfo{pages}{1} (\bibinfo{year}{1995}).

\bibitem[{\citenamefont{Rosa and Everaers}(2008)}]{Rosa2008}
\bibinfo{author}{\bibfnamefont{A.}~\bibnamefont{Rosa}} \bibnamefont{and}
  \bibinfo{author}{\bibfnamefont{R.}~\bibnamefont{Everaers}},
  \bibinfo{journal}{PLoS Comp. Biol.} \textbf{\bibinfo{volume}{4}},
  \bibinfo{pages}{1} (\bibinfo{year}{2008}).

\end{thebibliography}


\begin{thebibliography}{99}
\bibitem{resetting}
M.~R. Evans and S.~N. Majumdar.  Diffusion with Stochastic Resetting, {\it Phys. Rev. Lett.} {\bf 106}, 160601 (2011).
\bibitem{lammps} S. Plimpton. Fast Parallel Algorithms for Short-Range Molecular Dynamics. {\it J. Comput. Phys.} {\bf 117}, 1 (1995).
\bibitem{NAR} C.~A. Brackley, J. Johnson, S. Kelly, P.~R. Cook and D. Marenduzzo. Simulated binding of transcription factors to active and inactive regions folds human chromosomes into loops, rosettes and topological domains. {\it Nucleic Acics Res.} {\bf 8}, 3503 (2016).
\bibitem{langowski} J. Langowski. Polymer models of DNA and chromatin. {\it Eur. Phys. J. E} {\bf 19}, 241 (2006). 
\bibitem{Kremer1990} K. Kremer and G.~S. Grest. Dynamics of entangled linear polymer melts: A molecular-dynamics simulation. {\it J. Chem. Phys.} 92, 5057 (1990).
\bibitem{yeast} G.~G. Cabal, A. Genovesio, S. Rodriguez-Navarro, C. Zimmer, O. Gadal, A. Lesne, H. Buc, F. Feuerbach-Fournier, J.~C. Olivo-Marin, E.~C. Hurt and U. Nehrbass. SAGA interacting factors confine sub-diffusion of transcribed genes to the nuclear envelope. {\it Nature} {\bf 441}, 770 (2006).
\bibitem{Rosa2008} A. Rosa and R. Everaers. Structure and Dynamics of Interphase Chromosomes. {\it PLOS Comp. Biol.}, e1000153 (2008).
\bibitem{Oti2016}
M. Oti, J. Falck, M.~A. Huynen and H. and Zhou. CTCF-mediated chromatin loops enclose inducible gene regulatory domains. {\it BMC Genomics} {\bf 17}, 252 (2016).
\bibitem{angelo} A. Rosa, N.~B. Becker and R. Everaers. Looping Probabilities in Model Interphase Chromosomes. {\it Biophys. J.} {\bf 98}, 2410 (2010).
\bibitem{Shimada} J. Shimada and H. Yamakawa. Ring-closure probabilities for twisted wormlike chains. Application to DNA. {\it Macromolecules} {\bf 17}, 689 (1984).
\bibitem{Tang2015}
Z. Tang {\it et al.}. CTCF-mediated human 3D genome architecture reveals chromatin topology for transcription ,{\it Cell} {\bf 163}, 1611 (2015).
\bibitem{Sanborn2015} A.~L. Sanborn {\it et al.} Chromatin extrusion explains key features of loop and domain formation in wild-type and engineered genomes. {\it Proc. Natl. Acad. Sci. USA} {\bf 112}, E6456 (2015). 
\end{thebibliography}

\newpage

\renewcommand{\thefigure}{S\arabic{figure}}
\setcounter{figure}{0}

\appendix

\vspace{1cm}

\section{Exactly solvable non-equilibrium 1D models}

In this section we discuss the derivation of the solution of the 1D non-equilibrium model reported in the main text.

Let us denote by $p(x,t)$ the probability that the two monomers (heads) of a slip-link dimer (cohesin) are at a separation $x$, at a time $t$ after the slip-link binds to the chromatin fiber. The probability distribution $p(x,t)$ obeys the following generalized Fokker-Planck equation,
\begin{equation}\label{FPE}
\frac{\partial p(x,t)}{\partial t}=-k_{\rm off}p(x,t)+\frac{\partial}{\partial x}\left[\frac{1}{\gamma}\frac{dV}{dx}p(x,t)\right]+D\frac{\partial^2}{\partial x^2}p(x,t),
\end{equation}
where $k_{\rm off}$ is the slip-link detachment rate, $V(x)$ is the potential energy associated with the configuration in which the slip-link monomers hold a chromatin loop of size $x$, while $D$ and $\gamma$ are respectively the diffusion and drag coefficient for slip-links moving along the chromatin fiber. As usual, $D$ and $\gamma$ are related through the fluctuation-dissipation theorem (Stokes-Einstein formula), $D=k_BT/\gamma$. 
Eq.~(\ref{FPE}) should be solved with the initial condition that $p(x,t=0)=\delta(x-\sigma_{\rm sl})$, as we assume the slip-link binds to two adjacent regions of the chromatin fiber. Also, there is a reflecting boundary at $x=\sigma_{\rm sl}$, and an absorbing boundary at $x=l$: this is because once the slip-link binds to the convergent CTCF sites we assume that it ``clicks'' and sticks to them irreversibly (i.e. we assume that a pair of cohesin rings forming a bridge between two CTCF sites is a very stable complex).

The instantaneous probability at time $t$ that a slip-link with separation $\sigma_l<x<l$ unbinds from the chromatin fiber is
\begin{equation}
p_{\rm off}(t)dt= k_{\rm off} dt \int_{\sigma_{\rm sl}}^l dx \, p(x,t).
\end{equation}
In our simple analytical model, once the slip-link detaches, it cannot bind again; i.e., this is an absorbing state. Therefore the probability that the slip-link unbinds before reaching the $x=l$ absorbing state can be  found by integrating over all time,
\begin{equation}
P_{\rm off}=k_{\rm off} \int_0^{\infty} dt \, \int_{\sigma_{\rm sl}}^l dx \, p(x,t).
\end{equation}
As this is the probability that the slip-link unbinds while its separation is less than $l$, and as $x=l$ is an absorbing state, the probability that the system reaches the $x=l$ absorbing state is  given by
\begin{equation}
p(l)=1-P_{\rm off}.
\end{equation}
In other words, the slip-link cannot diffuse indefinitely in a finite 1D segment without either unbinding or reaching the absorbing state at $x=l$.

In order to solve these equations it is useful to define the following quantity, 
\begin{equation}
Q(x)=\int_0^{\infty} dt \, p(x,t).
\end{equation}
Note that $Q(x)$ may be viewed as the Laplace transform of $p(x,t)$,
$\hat{p}(x,s)$
\begin{equation}
\hat{p}(x,s)=\int_0^{\infty} dt \, e^{-st} p(x,t)
\end{equation} 
computed at $s=0$. By integrating Eq.~(\ref{FPE}) over time from $t=0$~to~$\infty$, we find that $Q(x)$ obeys the following ordinary differential equation
\begin{equation}\label{QODE}
-\delta(x-\sigma_{\rm sl})=-k_{\rm off}Q(x)+D\frac{d^2Q}{dx^2}+\frac{d}{dx}\left[\frac{1}{\gamma}\frac{dV}{dx}Q(x)\right],
\end{equation}
where the Dirac-delta function comes from the $t=0$ boundary condition. Since $P_{\rm off}$ has an absorbing boundary at $x=l$, we also have the boundary condition that $Q(l)=0$. The probability of eventually falling off the chromatin fiber is therefore $P_{\rm off}=k_{\rm off} \int_{\sigma_{\rm sl}}^l dx \, Q(x)$. 
Therefore, the probability of forming a CTCF-mediated loop is  equal to 
\begin{equation}
p(l)=1-k_{\rm off}\int_{\sigma_{\rm sl}}^l dx \, Q(x).
\end{equation}

Let us now compute $p(l)$ for the three cases discussed in the main text. For simplicity, in the active loop extrusion and free-diffusing cohesin model, we will consider from the start the limit $\sigma_{\rm sl}\to 0$, as a non-zero value of $\sigma_{\rm sl}$ is only required for the slip-link model with logarithmic potential. 

\subsection{Active extrusion model}

For the active extrusion model $D=0$ and $(1/\gamma)(dV/dx)=v$, where $v$ is the constant extrusion speed, so Eq.~(\ref{QODE}) reduces to a  first order differential equation in $Q(x)$,
\begin{equation}\label{extrusionQODE}
-\delta(x)=-k_{\rm off}Q(x)+vQ'(x),
\end{equation}
where $'$ denotes derivative with respect to $x$. The solution is
\begin{equation}
Q(x) = \frac{1}{v}e^{-k_{\rm off}x/v},
\end{equation}
so that 
\begin{equation}
p(l)=e^{-k_{\rm off}x/v}.
\end{equation}
Note that in this case we cannot apply the boundary condition at $x=l$ as the equation is first order.


\subsection{Free diffusing slip-link  model}

In the free diffusion model $dV/dx=0$, and the equation for $Q(x)$ is 
\begin{equation}\label{diffusionQODE}
-\delta(x)=-k_{\rm off}Q(x)+D Q''(x).
\end{equation}
The solution in this case is
\begin{equation}
Q(x) = \frac{1}{\sqrt{D k_{\rm off}}}\frac{e^{-\alpha x}-e^{-2\alpha l}e^{\alpha x}}{1+e^{-2\alpha l}},
\end{equation}
where here and in what follows we have defined $\alpha=\sqrt{k_{\rm off}/D}$, as in the main text. Consequently, the probability of forming a CTCF-mediated loop can be found to be
\begin{eqnarray}
p(l) 
& = & 1 - \frac{1-2e^{-\alpha l}+e^{-2\alpha l}}{1+e^{-2\alpha l}},  \\ \nonumber
& = & \frac{1}{{\rm cosh}(\alpha l)}.
\end{eqnarray}

\subsection{Slip-link in a logarithmic potential}

If the diffusing slip-link is subject to a logarithmic potential, $V(x)=ck_BT \log{x}$, which captures the entropic cost of looping, then the equation for $Q(x)$ is
\begin{equation}\label{sliplinkQODE}
-\delta(x-\sigma_{\rm sl})=-k_{\rm off}Q(x)+\left[\frac{a Q(x)}{x}\right]'+DQ''(x),
\end{equation}
where we have defined $a=ck_BT/\gamma$. The homogeneous version of Eq.~(\ref{sliplinkQODE}) can be written as
\begin{equation}\label{sliplinkQODE2}
DQ''(x)+\frac{aQ'}{x}-\frac{aQ}{x^2}-k_{\rm off}Q=0.
\end{equation}
If we write $Q=x^n f$, then $Q$ is a solution to Eq.~(\ref{sliplinkQODE2}) when the function $f(x)$ solves the following differential equation:
\begin{equation}\label{fODE}
x^2 f''+\left(2n+\frac{a}{D}\right)xf' -
\left[\frac{k_{\rm off}}{D}x^2 - (n-1)\left(n+\frac{a}{D}\right)\right]f=0.
\end{equation}
Now, by setting $n=(1/2)-(a/2D)=(1-c)/2$, we note that Eq.~(\ref{fODE}) can be written in the form,
\begin{equation}
x^2 y''+xy'-(x^2-A^2)y=0,
\end{equation}
where $A$ is a constant: this is the modified Bessel equation. Therefore the general solution of Eq.~(\ref{fODE}) can be written in terms of the modified Bessel functions of the first and second kind as follows,
\begin{equation}\label{fsolution}
f(x)=C_1 I_{m}(\alpha x)+C_2 K_{m}(\alpha x), \quad m=\frac{1}{2}+\frac{a}{2D},
\end{equation}
where $C_1$ and $C_2$ are constants to be determined from the boundary conditions, $I_m$ and $K_m$ respectively denote the order $m$ modified Bessel functions of the first and second kind. Note that $n+m=1$. 

To solve Eq.~(\ref{sliplinkQODE}), we note that it is equivalent to Eq.~(\ref{sliplinkQODE2}) with the following boundary conditions:
\begin{eqnarray}\label{sliplinkBC}
Q(l) & = & 0 \\ \nonumber
\left[DQ'+\frac{aQ}{x}\right]_{x=\sigma_{\rm sl}} & = & -1,
\end{eqnarray}
where the second boundary condition comes from integrating Eq.~(\ref{sliplinkQODE}) over an infinitesimal interval containing $x=\sigma_{\rm sl}$. Eqs.~(\ref{sliplinkBC}) can be used to determine the two constants $C_1$ and $C_2$ in Eq.~(\ref{fsolution}), to obtain
\begin{eqnarray}\label{C1C2}
C1 & = & - \frac{1}{D\alpha \sigma_{\rm sl}^n}\left[\frac{K_m(\alpha l)}{I_{m-1}(\alpha\sigma_{\rm sl})K_m(\alpha l)+K_{m-1}(\alpha \sigma_{\rm sl})I_m(\alpha l)}\right],\\ 
C2 & = & \frac{1}{D\alpha \sigma_{\rm sl}^n}\left[\frac{I_m(\alpha l)}{I_{m-1}(\alpha\sigma_{\rm sl})K_m(\alpha l)+K_{m-1}(\alpha \sigma_{\rm sl})I_m(\alpha l)}\right].
\end{eqnarray}
Note that we have used the following identities:
\begin{eqnarray}
I'_m(\alpha x) & = & I_{m-1}(\alpha x)-\frac{m}{\alpha x}I_m(\alpha x), \\ 
\nonumber
K'_m(\alpha x) & = & -K_{m-1}(\alpha x)-\frac{m}{\alpha x}K_m(\alpha x).
\end{eqnarray}
The solution of Eq.~(\ref{sliplinkQODE}) which satisfies the relevant boundary conditions is then given by
\begin{equation}\label{sliplinkQsol}
Q(x) = \left(\frac{x}{\sigma_{\rm sl}}\right)^n \frac{1}{D\alpha}
\left[\frac{I_m(\alpha l)K_m(\alpha x)-I_m(\alpha x) K_m(\alpha l)}{I_{m-1}(\alpha\sigma_{\rm sl})K_m(\alpha l)+K_{m-1}(\alpha\sigma_{\rm sl})I_m(\alpha l)}\right]. 
\end{equation}
From this we obtain
\begin{eqnarray}\label{sliplinkpl}
p(l) & = & 1-k_{\rm off}\int_{\sigma_{\rm sl}}^l dx \, Q(x) \\ \nonumber
 & = & \left(\frac{l}{\sigma_{\rm sl}}\right)^n
\frac{I_{m-1}(\alpha l)K_m(\alpha l)+I_m(\alpha l)K_{m-1}(\alpha l)}{I_{m-1}(\alpha \sigma_{\rm sl})K_m(\alpha l)+I_m(\alpha l)K_{m-1}(\alpha\sigma_{\rm sl})},
\end{eqnarray}
where we have used the following identities:
\begin{eqnarray}
\int dx \, x^n I_m(\alpha x) & = &  \frac{x^nI_{m-1}(\alpha x)}{\alpha} \\
\nonumber
\int dx \, x^n K_m(\alpha x) & = & -\frac{x^nK_{m-1}(\alpha x)}{\alpha},
\end{eqnarray}
which hold for indefinite integrals provided that $n+m=1$.

\section{Non-equilibrium 1D models of a single slip-link: stochastic simulations}

In this section we consider 1D stochastic simulations of a single slip-link diffusing in a logarithmic potential in the presence of two CTCF proteins, at mutual distance $l$, which act as barriers. For simplicity, as in the main text (Fig.~1) we only consider the relative distance between the slip-link monomers, $x$, and assume that it performs a random walk in an effective potential, whereas in reality both monomers diffuse and are subject to a potential dependent on the monomer-monomer separation -- we expect the two situations to be qualitatively analogous. With respect to the case considered in the main text, we here assume that there are no absorbing states, but rather that the slip-link gains an energy $\epsilon$ when it reaches a separation between the monomers $x=l$ (i.e., sticking between CTCF and cohesin is not permanent here, so $x$ can decrease later on). Correspondingly, the detachment rate will decrease at $x=l$: for concreteness, we assume $k_{\rm off}$ is constant, and equal to $k_0$, for $x\ne l$, while it is equal to $k_{\rm off}=k_0e^{-\epsilon/(k_BT)}$ for $x=l$. The single cohesin we model, once off, rebinds at rate $k_{\rm on}=k_{0}$, and when it does the monomers always start close together, so $x=\sigma_{\rm sl}$ (which is equal to the lattice spacing in our simulations). The logarithmic potential is $V(x)=c k_BT \log{x}$, and we choose here $c=2.1$ which corresponds to the formation of internal loops in a self-avoiding walk (see discussion in the main text, different values of $c$ lead to the same qualitative trends). The logarithmic potential and CTCF-cohesin interactions are incorporated in the algorithm via a standard Metropolis acceptance test.

\begin{figure}[!h]
	\centering
        \includegraphics[width=0.45\textwidth]{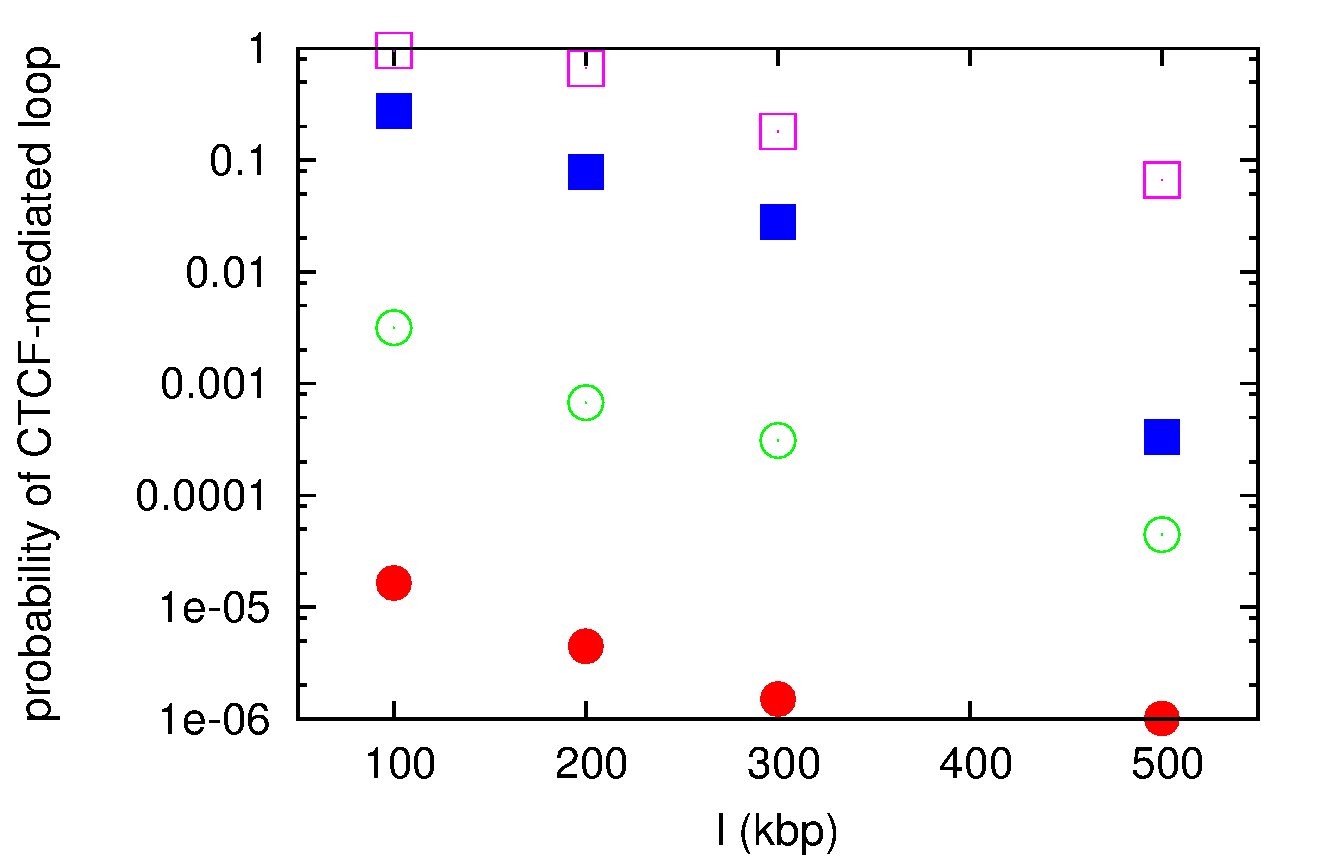}
	\caption{Plot of the probability of formation of a CTCF-mediated loop as a function of loop size, for different values of $\epsilon$. From bottom to top, curves correspond to $\epsilon=0$, $5 k_BT$, $8 k_BT$, $10 k_BT$ and $15 k_BT$ respectively. The case of $\epsilon=0$ models a divergent CTCF loop, the other cases with $\epsilon\ne 0$ model convergent CTCF loops with different assumptions for the strength of the thermodynamic attraction between CTCF and the slip-link.}
	\label{figS1}
\end{figure}

Fig.~\ref{figS1} shows a plot of the probability that the slip-link is on and has $x=l$ (i.e., the probability that a CTCF-mediated loop forms) once steady state is reached. As might be expected, we find that increasing $\epsilon$ strongly favours the CTCF-mediated loops, with respect to other states where the slip-link subtends a smaller loop size. This case is instructive because it suggests that a thermodynamic directional attraction between CTCF and cohesin (here, the interaction parametrised by $\epsilon$) is sufficient to favour the formation of CTCF-mediated loops. It should be noted that the model is still a non-equilibrium one, because $k_{\rm off}$ is constant for $x\ne l$, and, mainly, because upon rebinding the slip-link always returns to the case with $x=\sigma_{\rm sl}$. This second feature renders our model (both here and in the main text) to some extent similar diffusion ``with resetting'' model considered in~\cite{resetting}, although here the motion is further constrained by the logarithmic potential. Based on our results, we therefore suggest that non-equilibrium (re)binding (i.e., the resetting) and thermodynamic directional attraction are enough to explain the bias favouring the formation of convergent CTCF loops ($\epsilon\ne 0$) with respect to divergent ones (where there is no directional attraction, and hence $\epsilon=0$). Again, and as in the main text, because this is a non-equilibrium model, the probability of formation of CTCF-loop is not compatible with a power law: rather it decays approximately exponentially (see the log-linear plot in Fig.~\ref{figS1}).

\section{1D models with many interacting slip-links, and the osmotic ratchet}

\subsection{1D model without looping weight}

We now consider the case of multiple slip-links studied in the main text, and derive the formula for the density and effective extrusion force in the case where slip-links always rebind at the same ``loading site''. This is the case which leads to the osmotic ratchet discussed in the main text. In this section, we consider 1D models (3D simulations are described separately below).

\begin{figure}[!h]
\includegraphics[width=0.6\textwidth]{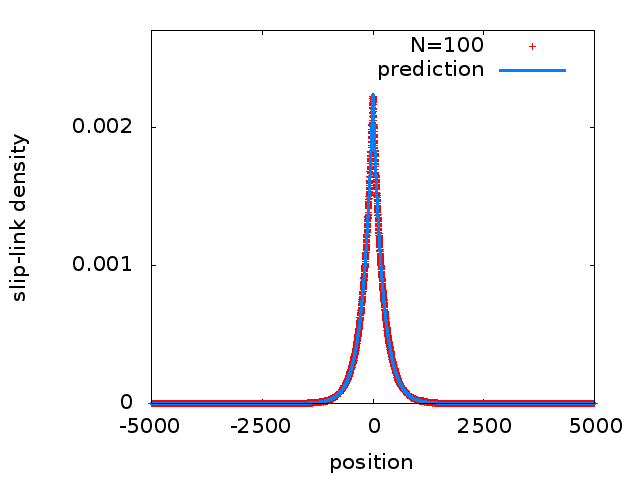}
\caption{Plot of the density of slip-link monomers as a function of position, for a chromatin fiber of length $L=10000$, with a loading site in the middle, at $x=0$, and $N=100$ slip-links which have $k_{\rm on}=k_{\rm off}=0.00001$, and $D=0.5$ (all parameters in simulation units). \label{figsldensity}}
\end{figure}

We first consider a simplified model without ``looping weights'', where $N$ slip-links simply diffuse on a chromatin fiber of length $L$: i.e., this model neglects the entropic cost associated with the formation of a given loop network. If we disregard excluded volume interactions we can write down the following partial differential equation for the (average) density $\rho(x,t)$ of slip-links bound to chromatin at position $x$, where the loading site is located at $x=0$,
\begin{eqnarray}\label{rhoPDE}
\frac{\partial \rho(x,t)}{\partial t} & = & k_{\rm on} N_{\rm off} \delta(x) -k_{\rm off}\rho(x,t) +D\frac{\partial^2 \rho(x,t)}{\partial x^2} \\ \nonumber
& = & \frac{k_{\rm on} k_{\rm off} N}{k_{\rm on}+k_{\rm off}} \delta(x) -k_{\rm off}\rho(x,t) +D\frac{\partial^2 \rho(x,t)}{\partial x^2},
\end{eqnarray}
where $N_{\rm off}=\frac{k_{\rm off}N}{k_{\rm on}+k_{\rm off}}$ is the average number of unbound cohesins (which are available to bind at the loading site). The three terms on the right hand side of Eq.~(\ref{rhoPDE}) respectively denote binding at the loading site with rate $k_{\rm on}$, unbinding with rate $k_{\rm off}$ from any site, and diffusion. Note that here $D$ is the diffusion constant for a slip-link monomer moving along the chromatin fiber. This equation does not include noise, therefore it should be seen as a mean field theory, which predicts the average value of $\rho(x,t)$. The steady state solution of Eq.~(\ref{rhoPDE}) which decays for $x\to \infty$ (relevant for $L\to \infty$) is given by 
\begin{equation}
\rho(x)=A e^{-\alpha |x|}
\end{equation}
where $A$ is a constant and in a similar way to before we define $\alpha=\sqrt{k_{\rm off}/D}$.  Similarly to what was previously done in the section ``Exactly solvable non-equilibrium models'', the constant $A$ can be determined by integrating Eq.~(\ref{rhoPDE}) around $0$, from $x=-\epsilon$ to $x=+\epsilon$, and then sending $\epsilon\to 0$. This procedure leads to the requirement that 
\begin{equation}
A=\frac{1}{2D\alpha} \frac{k_{\rm on} k_{\rm off} N}{k_{\rm on}+k_{\rm off}},
\end{equation}
and therefore $\rho(x)$ in steady state is given by
\begin{equation}\label{rhoratchet}
\rho(x)=\frac{N}{2}\frac{ k_{\rm on} k_{\rm off}}{k_{\rm on}+k_{\rm off}}\frac{1}{D\alpha}e^{-\alpha |x|}.
\end{equation}
Computer simulations of $N$ slip-links diffusing \textit{with} excluded volume interactions on a chromatin fiber of size $L$ confirm that the average density profile of bound slip-links is an exponentially decaying function centred on the loading site, in good agreement with Eq.~(\ref{rhoratchet}) even for a large number of slip-links (Fig.~\ref{figsldensity}).

The 1D pressure exerted by the slip-link gas is equal to $p(x)=N k_BT \rho(x)$; for a given slip-link at position $x$, there will be a difference in the pressure on the inside and outside of each head of the link, resulting on an outward force. Since the size of one of the slip-link head is $\sigma_{\rm sl}$, we can estimate the force acting on a head at position $x$ as follows:
\begin{eqnarray}\label{force}
f (x) & \simeq & -k_BT \sigma_{\rm sl} \frac{\partial \rho}{\partial x}
\\ \nonumber
 & = & \frac{k_BT \sigma_{\rm sl}N k_{\rm on}k_{\rm off} }{2D(k_{\rm on}+k_{\rm off})} e^{-\alpha |x|}
\end{eqnarray}
If we now imagine a slip-link placed symmetrically around the loading site, so that its two heads are at positions $\pm x$, then the osmotic pressure will tend to increase the size of the loop $l=2x$. If we assume for simplicity, that the loop will remain symmetrical with respect to the loading site, we can write down the following equation for the effective extrusion velocity of the loop, $v=dl/dt$,
\begin{equation}\label{extrusionv}
\gamma \frac{dl}{dt}=\frac{k_BT \sigma_{\rm sl}N k_{\rm on} k_{\rm off}}{D(k_{\rm on}+k_{\rm off})} e^{-\alpha l/2},
\end{equation}
where $\gamma$ is the slip-link's effective drag coefficient. Eq.~(\ref{extrusionv}) predicts that the maximal extrusion speed is when the loop is close to the loading site, where it can be approximated as
\begin{equation}\label{maxextrusionv}
v\sim v(l\to0)=\frac{\sigma_{\rm sl}k_{\rm on}k_{\rm off}N}{(k_{\rm on}+k_{\rm off})},
\end{equation}
where we have used the fluctuation-dissipation relation $D=k_BT/\gamma$. Note that the solution of Eq.~(\ref{extrusionv}) is given by
\begin{equation}\label{lt}
l(t)=\frac{2}{\alpha}\log\left[1+\frac{N k_{\rm on}k_{\rm off}\alpha\sigma_{\rm sl}}{2(k_{\rm on}+k_{\rm off})} t\right],
\end{equation}
so that this simple theory predicts that extrusion should slow down with loop size, which should only increase logarithmically at later times. Note that Eq.~(\ref{lt}) predicts the average evolution of the loop size for a slip-link binding at the loading site at $t=0$, whereas in Fig.~\ref{figsldensity} in the main text we plot the size and distribution probability of the {\it largest} loop at a given time. However, this simplified theory is useful as it clarifies that loops can be extruded provided the steady state slip-link density $\rho(x)$ is not constant. Of course, if there is not a preferred loading site, the first term in Eq.~(\ref{rhoPDE}) becomes $k_{\rm on} N_{\rm off}/L$: in this case $\rho(x)$ is constant in steady state, and there is no longer an osmotic pressure driving extrusion, in line with the results discussed in the main text for the case with random rebinding. 

\begin{figure}[!h]
	\centering
        \includegraphics[width=0.7\textwidth]{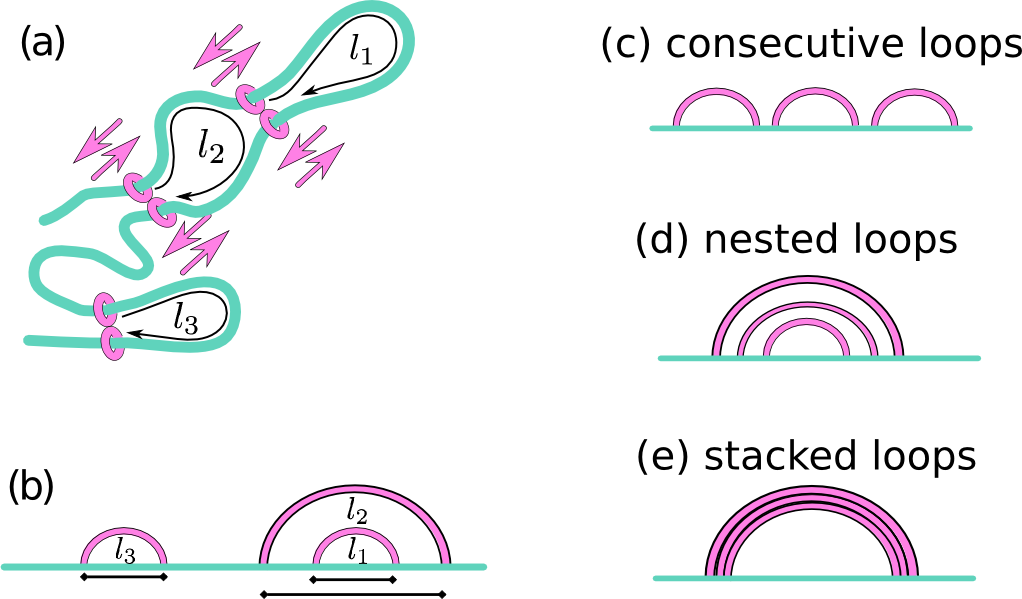}
	\caption{(a,b) Cartoon of a network of slip-links, and associated loops, viewed either in 3D (a), or 1D (b). In this network, there are three loops: $l_1$, $l_2$ and $l_3$. Of these, $l_1$ and $l_3$ are ``simple loops''. (c-e) Cartoons of looping diagrams which typically arise in our simulations. The arrangement of slip-links in (c) is referred to as ``consecutive loops'', while that in (d) is referred to as ``nested loops''. For nested loops, if arcs are close together, this corresponds to a 3D structure where slip-links are closely stacked onto one another (see also Fig.~4 in the main text): we referred to this pattern as ``stacked loops''. If different arcs are coloured differently, stacked arcs appear as rainbow patterns (Suppl. Movies 2 and 3).}
	\label{figloopnetwork}
\end{figure}

\subsection{1D Model with looping weight, and looping diagrams}

The model discussed above corresponds to the case ``without looping weight''. The case ``with looping weight'' discussed in the main manuscript can be considered by introducing an entropic potential which affects the motion of slip-link monomers (in practice, this is done through a standard Metropolis test). For simplicity, we assume that all loops are Gaussian, i.e., we disregard self-avoidance effects in this calculation. To compute the looping of a given configuration of slip-link heads (e.g., that in Figs.~\ref{figloopnetwork}a,b), we first identify all loops. The number of loops, $n$, is equal to the number of bound slip-links, $N_b$, and we label their sizes $l_1,\ldots,l_{n}$ (see Figs.~\ref{figloopnetwork}a,b). We then identify the number of ``simple loops'', which do not contain another loop inside. In general, there will be a number $n_s\leq n$ of simple loops. The probability of formation of each loop is $\sim l^{-3/2}$, and this is weighted by another factor $e^{-\tilde{\kappa}/l^2}$ for simple loops to model the energetic cost of bending; $\tilde{kappa}$ is a constant associated with the persistence length of the chromatin fiber. The looping weight is then
\begin{equation}
w_{\rm looping}=\left(\Pi_{i=1,\ldots,n}\frac{1}{l_i^{3/2}}\right) \left(\Pi_{i=1,\ldots,n_s}e^{-\frac{\tilde{\kappa}}{l_i^2}}\right). 
\end{equation}
This looping weight is defined up to a multiplicative constant, and, in turn, it defines the potential in which the slip-links move (up to an irrelevant additive constant) via
\begin{equation}
V_{\rm looping}=-k_BT \log{w_{\rm looping}}. 
\end{equation}
In Fig.~3 in the main manuscript we present results without looping weight; Fig.~\ref{ratchet} shows the results of simulations with looping weight, with $\kappa=8$ (in units of $\Delta x^{-2}$, where $\Delta x=\sigma_{\rm sm}/2$), a choice corresponding to a rather flexible polymer. The results show that the looping weight makes a notable quantitative change, but the qualitative trends are very similar to those in Fig.~3 in the main manuscript, with the model with loading leading to the ratchet effect discussed above and in the main manuscript. 

Diagrams such as that in Figure~\ref{figloopnetwork} are useful to determine visually the looping topology without the need to show the 3D configuration. Such diagrams are used extensively for RNA secondary structure representations; we refer to these in our context as ``looping diagrams''. In the text we refer to some specific loop configurations which are most easily described by these diagrams: these are the ``consecutive loop'' arrangement in Figure~\ref{figloopnetwork}c, the ``nester loop'' one in Figure~\ref{figloopnetwork}d, and ``the stacked loops'' of Figure~\ref{figloopnetwork}e, where some of the loops in a nested loop are packed close to each other. As discussed in the main text, stacked loops are entropically favoured, hence they appear often in our simulations: in Supplementary Movies 2 and 3, where each arc is colored differently, stacked loops appear as rainbow patterns.

\begin{figure}[!h]
\renewcommand{\thesubfigure}{a}
\subfloat[]{\includegraphics[width = .49\textwidth]{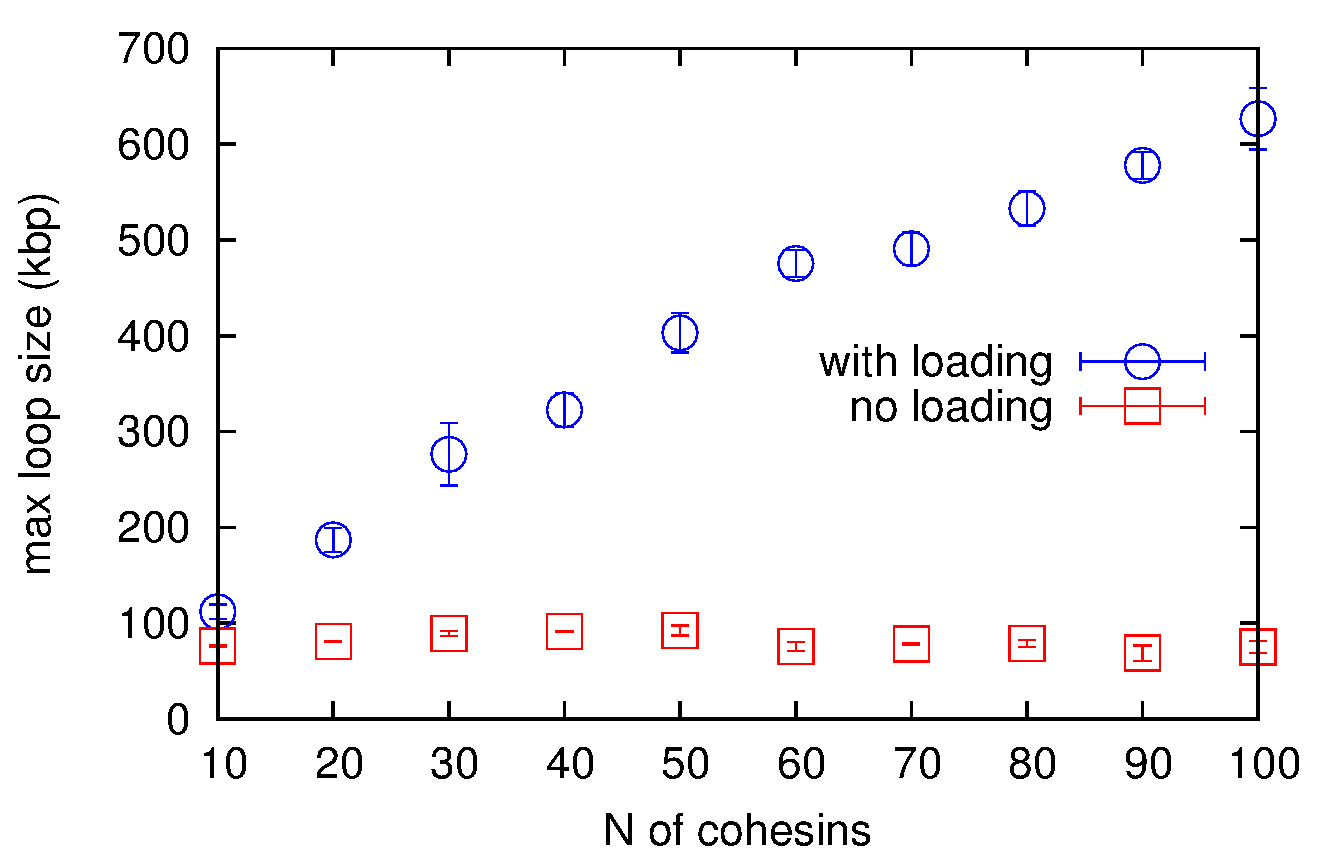}}
\renewcommand{\thesubfigure}{b}
\subfloat[]{\includegraphics[width = .49\textwidth]{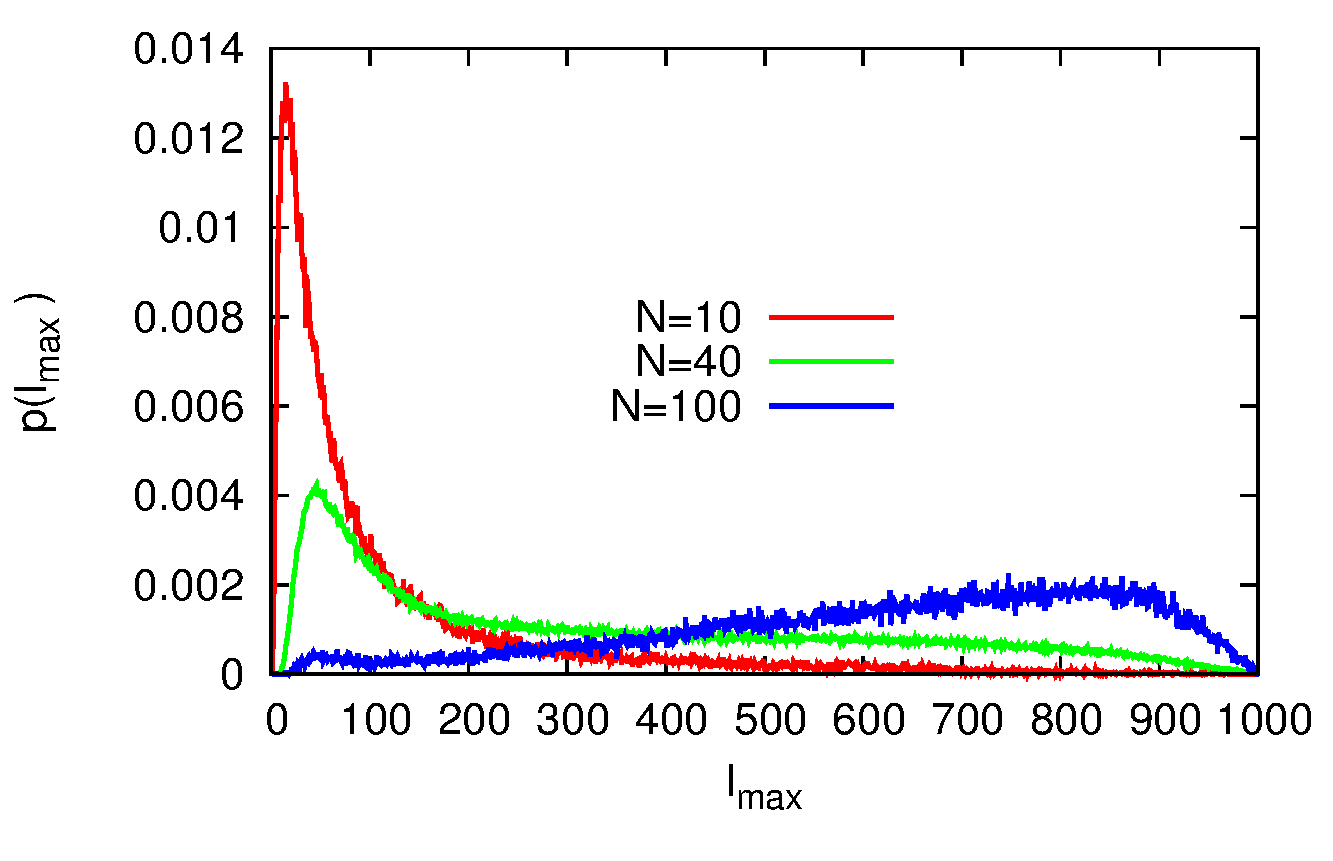}}
\caption{The osmotic ratchet in 1D simulations with looping weight. (a) Plot of the time average of the largest loop, $\langle l_{\rm max}\rangle$, as a function of $N$ for the model with and without loading (see legend). (b) Distributions of the largest loop size for different $N$, for the model with loading.}
	\label{ratchet}
\end{figure}


\section{3D Brownian dynamics of a chromatin fiber with a single molecular slip-links}

In this section we give details and additional results for the three-dimensional Brownian dynamics simulations of a slip-link sliding diffusively on a chromatin fiber, which are discussed in the main text.

\subsection{Brownian dynamics: force field and other simulation details}

In our Brownian dynamics simulations we follow the evolution of a chromatin fiber and of a slip-link which is topologically bound to the fiber. The dynamics are evolved using a velocity-Verlet integration scheme within the LAMMPS software~\cite{lammps} in Brownian dynamics mode (NVT ensemble). 

The chromatin fiber is modelled, as in Ref.~\cite{NAR}, as a bead-spring self-avoiding and semi-flexible polymer; each of its beads have size $\sigma$.
If we denote the position of the centre of the {$i$}-th chromatin bead by {$\mathbf{r}_i$}, and the separation between beads $i$ and $j$ by $d_{i,j}=|\mathbf{r}_i-\mathbf{r}_j|$, then we can express the finitely-extensible non-linear (FENE) spring potential modelling the connectivity of the chain as follows:
\begin{equation}\label{Ufene}
U_{\rm FENE}(i,i+1) = -\dfrac{k}{2} R_0^2 \ln \left[ 1 - \left( \dfrac{d_{i,i+1}}{R_0}\right)^2\right],  
\end{equation}
for {$d_{i,i+1} < R_0$} and  {$U_{\rm FENE}(i,i+1) = \infty$}, otherwise; here we chose  {$R_0 = 1.6$ $\sigma$} and  {$k=30$}  {$\epsilon/\sigma^2$}. 

The semi-flexibility (bending rigidity) of the chain is described through a standard Kratky-Porod potential, defined in terms of the positions of a triplet of neighbouring beads along the polymer as follows:
\begin{equation}\label{Ubend}
U_{\rm B}(i,i+1,i+2) = \dfrac{k_BT l_p}{\sigma}\left[ 1 - \dfrac{\bm{d}_{i,i+1} \cdot \bm{d}_{i+1,i+2}}{d_{i,i+1}d_{i+1,i+2}} \right],
\end{equation}
where we set the persistence length $l_p = 4 \sigma$ {(which maps to  $\simeq 120$~nm -- see below; this is reasonable for chromatin~\cite{langowski}). 

Self-avoidance is ensured by introducing a repulsive
Weeks-Chandler-Anderson (WCA) potential between every chromatin bead as follows:
\begin{equation}\label{WCA}
U_{\rm LJ}(i,j) = 4 \epsilon \left[ \left(\dfrac{\sigma_c}{d_{i,j}}\right)^{12} - \left(\dfrac{\sigma_c}{d_{i,j}}\right)^6 \right] +\epsilon, 
\end{equation} 
for $d_{i,j}< 2^{1/6}\sigma$, and $U_{\rm LJ}(i,j) =0$ otherwise. In Eq.~(\ref{WCA}) we set $\epsilon=k_BT$.

The total potential energy experienced by chromatin bead $i$ is given by
\begin{equation}
	U_i=\sum_j U_{\rm FENE}(i,j)\delta_{j,i+1} + \sum_j\sum_k U_{\rm B}(i,j,k)\delta_{j,i+1}\delta_{k,i+2} + \sum_j U_{\rm LJ}(i,j),
\end{equation}
and its dynamics can be described by the Langevin equation
\begin{equation}
m \ddot{\mathbf{r}}_i = - \xi \dot{\mathbf{r}}_i - {\nabla U_i} + \boldsymbol{\eta}_i,
\label{langevin}
\end{equation}
where $m$ is the bead mass, $\xi$ is the friction coefficient, and $\boldsymbol{\eta}_i$ is a stochastic delta-correlated noise. The variance of each Cartesian component of the noise, $\sigma_{\eta}^2$, satisfies the usual fluctuation dissipation relation $\sigma_{\eta}^2 = 2 \xi k_B T$.

\begin{figure}
\includegraphics{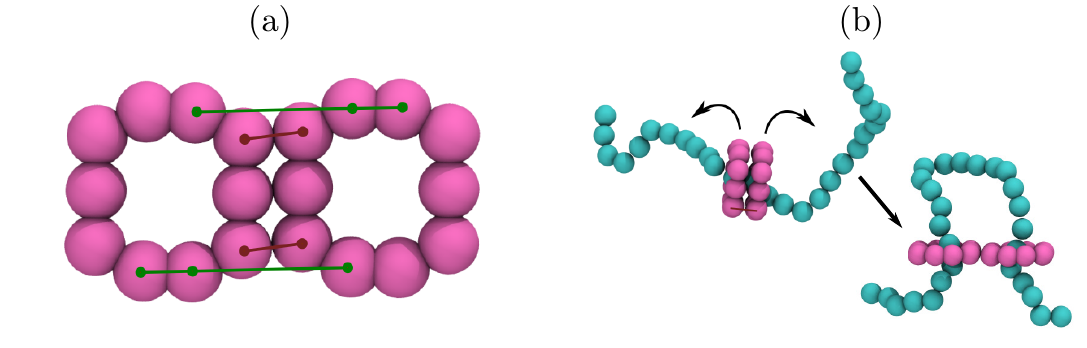}
\caption{\label{BDcohesinfig} Snapshots showing the slip-link model used in the Brownian dynamics simulations. (a) Each slip-link consists of a pair of rings composed from 10 beads of diameter $\sigma$. Each ring moves as a rigid body, and the pair is held together by two FENE bonds, indicated by brown lines. To keep the rings in an open handcuff configuration two Kratky-Porod bending interactions are added between beads as indicated by green lines. The separation between the centre point of each ring is $2R=3.5 \sigma$. (b) When a slip-link is added to the chromatin fiber, it is first arranged as a folded handcuff, with each ring encircling an adjacent bead of the polymer. As the dynamics evolve the bending potential acts to quickly unfold the handcuff and generate a loop.}
\end{figure}

In order to model slip-links we build a pair of rings out of beads also of diameter $\sigma$, and allow each ring to move as a rigid body. The translational motion of the centre of mass of the ring is described by a Langevin equation as in Eq. (\ref{langevin}), while rotation is described by a similar equation where the force term is replaced by the torque on the centre of mass, calculated from the forces experienced by the component beads of the ring. Each ring is composed of 10 beads arranged so that it is large enough to encircle the chromatin fiber. The two rings are held together by a pair of FENE bonds (as in Eq. (\ref{Ufene})), and they are kept in an open ``handcuff'' arrangement via two bending interactions (as in Eq. (\ref{Ubend}), but with $l_p=100\sigma$). The slip-link beads interact with each other, and with chromatin beads with the WCA potential described above. Figure~\ref{BDcohesinfig}A shows the arrangement of a pair of rings and indicates the interactions between them. Slip-links are attached to a chromatin fiber by first positioning them in a folded handcuff arrangement such that each ring encircles an adjacent polymer bead; the bending interactions between the two rings then act to open the the handcuff, and bend the polymer (see Fig.~\ref{BDcohesinfig}B). After this the slip-link is free to diffuse in 3D and along the polymer.

As is customary~\cite{Kremer1990}, we use simulation units where the mass of a polymer bead $m=1$, and the distance and energy units $\sigma=1$ and $\epsilon=k_BT$ respectively; the simulation time unit is given by $\tau_{\rm LJ} = \sigma \sqrt{m/\epsilon}$. There are two other time scales in the system, the velocity decorrelation time $\tau_{\rm in}=m/\xi$ and the Brownian time $\tau_{\rm B}=\sigma/D_b$; we set the friction $\xi=1$ meaning that $\tau_{\rm in}=\tau_{\rm LJ}=\tau_{\rm B}$. Here $D_b = k_BT/\xi$ is the diffusion coefficient of a bead of size $\sigma$.  From the Stokes' friction coefficient for spherical beads of diameter $\sigma$ we have that $\xi = 3 \pi \eta_{sol} \sigma$ where $\eta_{sol}$ is the solution viscosity. 
For the slip-link rings we set a total mass of each ring of $m_r=2.75m$; keeping $\tau_{\rm in}= \tau_{\rm LJ}$ ensures a suitably larger friction $\xi_r$ for these larger proteins (this means we approximate that each ring diffuses like a sphere of diameter 2.75$\sigma$). The numerical integration of Eq.~(\ref{langevin}) uses a time step $\Delta t = 0.01 \tau_{\rm LJ}$.

The mapping from simulation to physical units can be made as follows. Energies are mapped in a straightforward way as they are measured in units of $k_B T$. To map length scales from simulation to physical units, we set the diameter, $\sigma$, of each bead to $\sim 30$nm$\simeq 3$ kbp (assuming an underlying 30 nm fiber hence a $100$ bp/nm compaction; of course, all our results would remain valid with a different mapping). For time scales, by requiring that the mean square displacement of a polymer bead matches that measured experimentally in Ref.~\cite{yeast}, as done in Ref.~\cite{Rosa2008}, we obtain that $\tau_B=\tau_{\rm LJ}=0.1$ s.

\subsection{Brownian dynamics: additional results}

In Fig.~2 in the main text, we have shown the looping frequency for a slip-link, with a flexible chromatin fiber ($l_p=4\sigma$). While this value is appropriate for open chromatin, which is usually found within CTCF-mediated loops~\cite{Oti2016}, from a theoretical point of view it is of interest to ask what is the effect of changing the persistence length $l_p$.

\begin{figure}[!h]
\includegraphics{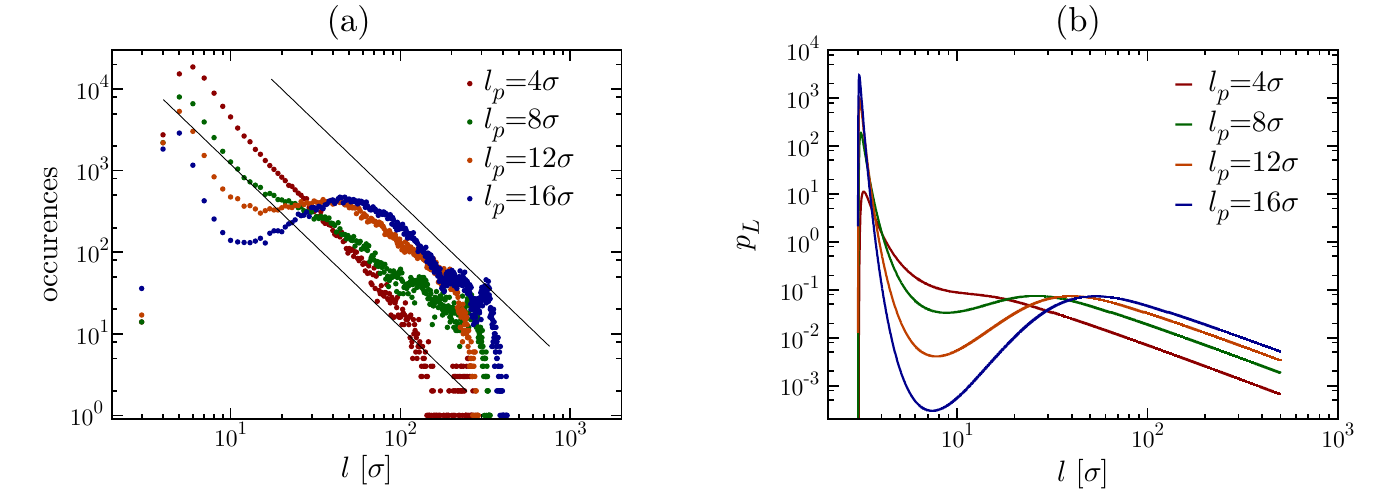}
\caption{(a) Log-log plot of the frequency of looping for slip-links diffusing on chromatin fibers with different flexibility (i.e., different values of the persistence length, $l_p$). These results are from Brownian dynamics simulations. (b) Probability of loop formation for an ideal semi-flexible polymer with different values of $l_p$ as a function of loop size, according to the analytical approximation in Ref.~\cite{angelo}, see the formula in the text. We assumed that the two ends of the loop need to lie at a 3D distance of $r=3\sigma$ from each other, and scaled the probability distributions by $8 l_p^3$ to show them more clearly on the same graph. The case described in the main text (Fig.~2b) considers the most relevant parameters set for chromatin with $\sigma=30$ nm and $l_p=4\sigma=120$ nm.   \label{figsimloops}}
\end{figure}

Figure~\ref{figsimloops}A shows the frequency of looping as a function of loop size, as found with our Brownian dynamics simulations. It can be seen that, while the first peak, for smaller values of the loop length, remains in a similar position for all values of $l_p$, for stiffer polymers there is a second shoulder, or smaller peak, for larger values of the loop length. Fig.~\ref{figsimloops}B shows a prediction of looping probabilities obtained by using the analytical estimate in Ref.~\cite{angelo}. These results show that both peaks can be explained by considering a simple theory for semi-flexible polymers which neglects excluded volume interactions, by assuming that looping via a slip-link is equivalent to the constraint that the two ends of a loop are separated in 3D by a distance $r$. Physically, the first peak arises because very small loops cannot form, as a loop must at least span a distance $r$. The second peak is related to the well known optimal size of a loop in a semi-flexible polymer, which comes about due to the competition between the entropic cost, which favours shorter loops, and bending penalties, which favours longer loops~\cite{Shimada}. 

For completeness, we report here the form of the analytical approximation for the distribution probability of the end-to-end distance $r$ of a semiflexible polymer of size $L$, $p_L(r)$, used in Ref.~\cite{angelo}:
\begin{eqnarray}\label{analyticalpLr}
p_L(r) = & J(L) & \left(\frac{1-c\rho^2}{1-\rho^2}\right)^{5/2}
\exp\left(\frac{\sum_{i=-1}^0\sum_{j=0}^3 c_{ij}\lambda^i \rho^{2j}}{1-\rho^2}\right)
\\ \nonumber
& & \exp\left(-\frac{d\lambda ab(1+b)\rho^2}{1-b^2\rho^2}\right) 
I_0\left(-\frac{d\lambda a(1+b)\rho^2}{1-b^2\rho^2}\right),
\end{eqnarray}
where
\begin{eqnarray}\label{constantspLr}
\rho & = & \frac{r}{L} \\ \nonumber
\lambda & = & \frac{l_p}{L} \\ \nonumber
a & = & 14.054 \\ \nonumber
b & = & 0.473 \\ \nonumber
c & = & 1-\left[1+\left(0.38\lambda^{-0.95}\right)^{-5}\right]^{-0.2} 
\\ \nonumber  
c_{ij} & = & \left( \begin{array}{ccc} 
-3/4 & 23/64 & -7/64 \\ \nonumber
-1/2 & 17/16 & -9/16 
\end{array}
\right) 
\\ \nonumber
1- d & = & \left\{ 
\begin{array}{cc}
0 & {\rm if} \, \lambda < 1/8 \\ \nonumber
\frac{1}{\frac{0.177}{\lambda-0.111}+6.4 \left(\lambda-0.111\right)^{0.783}} & {\rm if} \, \lambda \ge 1/8 
\end{array}
\right.
\\ \nonumber
J(L) & = & \frac{1}{8l_p^3}\left\{
\begin{array}{cc}
\left(\frac{3\lambda}{\pi}\right)^{3/2} \left[1-\frac{5\lambda}{4}-\frac{79\lambda^2}{160}\right]  & {\rm if} \, \lambda < 1/8 
\\ \nonumber
896.32 \lambda^5 \exp\left(-14.054 \lambda +\frac{0.246}{\lambda}\right) 
& {\rm if} \, \lambda \ge 1/8 
\end{array}
\right. .
\end{eqnarray}
In Eq.~(\ref{constantspLr}), $I_0$ is a modified Bessel function of the first kind, and $J(L)$ is the so-called Shimada and Yamakawa $J$-factor, measuring the ring closure probability for a wormlike chain (see~\cite{Shimada} for details and the derivation of this factor).

\begin{figure}
\includegraphics{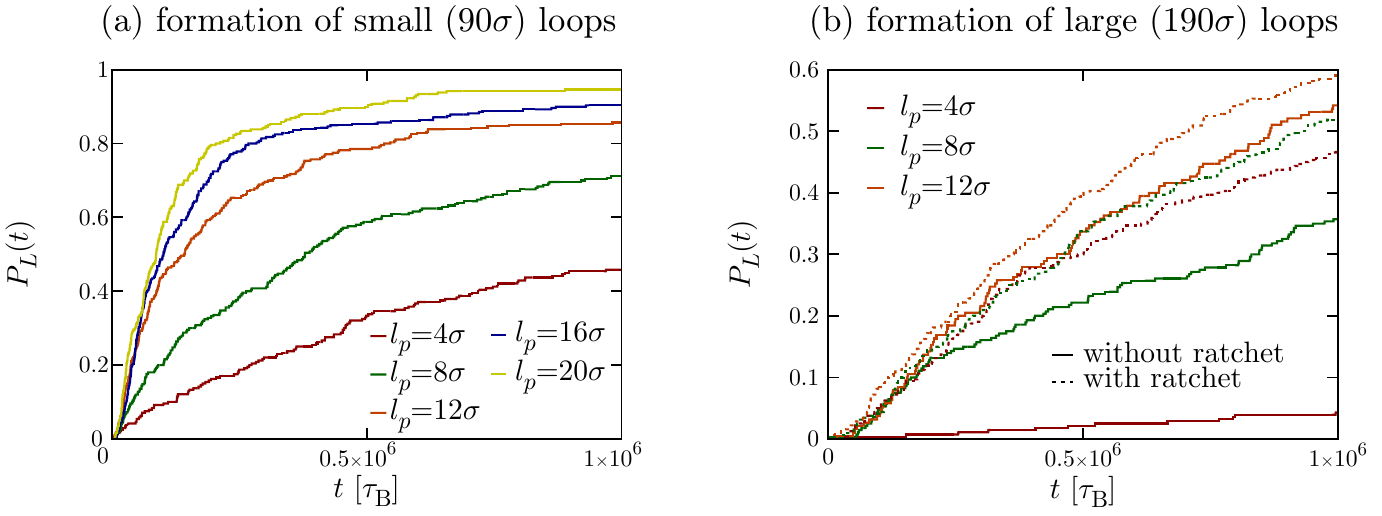}
\caption{\label{BDcohesinadditional} (a,b) Curves showing the fraction of loops of size $90\sigma$ (a) or $190\sigma$ (b) which form as a function of time (measured in timesteps, one Brownian time corresponds to $100$ timesteps). The simulations ``with ratchet effect'' contain three slip-links per CTCF loop (see text), the others one slip-link per CTCF loop. Using parameters relevant for chromatin ($\sigma=30$ nm with a compaction of $100$ bp/nm (i.e., $3$ kbp per bead), these results correspond the two loop sizes of $270$ kbp and $570$ kbp.}
\end{figure}

Finally, Fig.~\ref{BDcohesinadditional} shows results from simulations of a model chromatin fiber of size $L=2000\sigma$, split into sections, in each of which we place a slip-link (see Fig.~4 of the main manuscript). At the end of each section we locate beads with high affinity for the slip-link -- modelling convergent CTCF sites (the end ``CTCF bead'' of a given section is $5\sigma$ away from the start ``CTCF bead'' of the next section). The results show the fraction of slip-links which reach the sticky CTCF sites for different values of the persistence length, $l_p$, and of the loop, or section, size (see also Suppl. Movie 1, valid for $l_p=8\sigma$ and a loop size of $90\sigma$). The simulations ``with ratchet effect'' consider three slip-links per section, with the topology arranged so as to give three loops where the largest loop contains the middle loop which in turn contains the smallest loop (see also Suppl. Movie 2, and the corresponding nested rainbow rings determining looping topology). The ``ratchet'' effect, which is discussed more in detail in the next Section, leads to a dramatic increase in the fraction of large loops which can be formed within a flexible fiber. Our results also show that the stiffer the fiber, the more likely is the formation of a CTCF-mediated loops (Fig.~\ref{BDcohesinadditional}B). 

\subsection{3D Brownian dynamics with multiple slip-links}

In Figure~3d of the main text we present results from 3D Brownian dynamics simulations of multiple slip-links (see also Suppl. Movie 3). These simulations were performed with the same geometry and force field described in the Section ``3D Brownian dynamics of a chromatin fiber with a single molecular slip-links'', but now we consider $N$ slip-links, which can additionally: (i) bind with rate $k_{\rm on}$ if detached, (ii) detach with rate $k_{\rm off}$ if bound. In practice, to simulate this we have  coupled LAMMPS with an in-house code modelling stochastic detachment and binding. Detached slip-rings are not simulated directly via Brownian dynamics  but are taken into account to determine which slip-links unbind and which rebind. \\

\section{Analysis of ChIA-PET data}

To estimate the \textit{in vivo} probability of finding cohesin and CTCF-mediated loops of a given length, we analysed ChIA-PET (Chromatin Interaction Analysis by Paired-End Tag Sequencing) data from Ref.~\cite{Tang2015} (data publicly available from the Gene Expression Omnibus (GEO), accession number GSE72816). In these experiments chromatin-chromatin interactions mediated by a specific proteins are identified using immunoprecipitation; in this way pairs of interacting chromatin regions which are both bound by CTCF are identified. In particular, Fig.~1d of the main manuscript shows the contact probability between CTCF-bound sequences in GM12878 cells (data set from GEO accession number GSM1872886). Data were sorted into bins of size $5$ kbp according to loop size. This contact probability, in the range $100-1000$ kbp, is compatible with an exponential decay (see Fig. 1d), where the decay length of the exponential is of the order of hundreds of kbp, within the typical range of CTCF-mediated loops~\cite{Oti2016}. These data can also be fitted by a power law, but the fit is quite poor (Figs. 1e or~\ref{figChIAPET}a). The exponent resulting from the fit is $\sim -0.35$, which is lower than and far from those which can be explained by a polymer physics model (e.g., $\sim -2.1$ for an ideal self-avoiding walk, $\sim -1.5$ for an equilibrium globule or $\sim -1$ for a fractal globule). 
It is interesting to note that ChIA-PET contacts between sequences bound to RNA polymerase II appear to be better fitted by a power law albeit with a similarly low exponent (Fig.~\ref{figChIAPET}b; data set from GEO accession number GSM1872887~\cite{Tang2015}). These result suggest that CTCF-mediated contacts (as well as RNA PolII-mediated ones) appear to obey contact decay laws which are incompatible with the laws which would be predicted on the basis of equilibrium polymer models. Notably, such polymer models can however account very well for the contact decay laws typically found in Hi-C experiments~\cite{Sanborn2015} (which probe chromatin-chromatin interactions genome wide without selecting for specific proteins). A possible explanation is that Hi-C contacts encompass many interactions arising randomly from spatial proximity in 3D (i.e. there does not have to be a protein mediated interaction, since any chromatin regions with close proximity can be captured), which can be explained by polymer physics assuming randomly diffusing polymers. Taken together these observations support our hypothesis that cohesin/CTCF mediated loops form through a non-equilibrium mechanism \textit{in vivo}, and are distinct from other chromatin diffusion mediated loops. 

\begin{figure}[!h]
\renewcommand{\thesubfigure}{a}
\subfloat[]{\includegraphics[width = .49\textwidth]{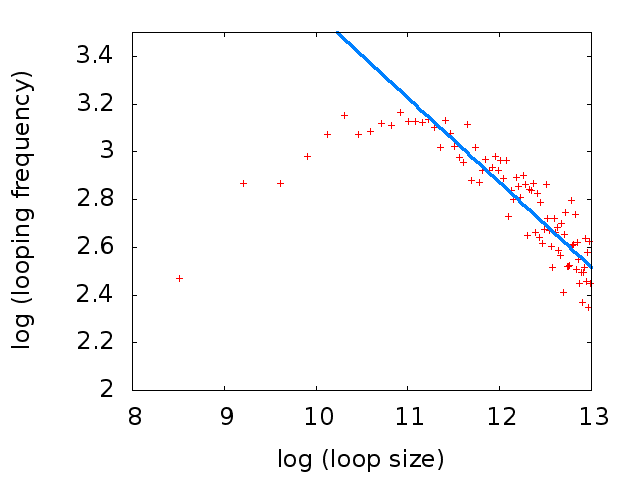}}
\renewcommand{\thesubfigure}{b}
\subfloat[]{\includegraphics[width = .49\textwidth]{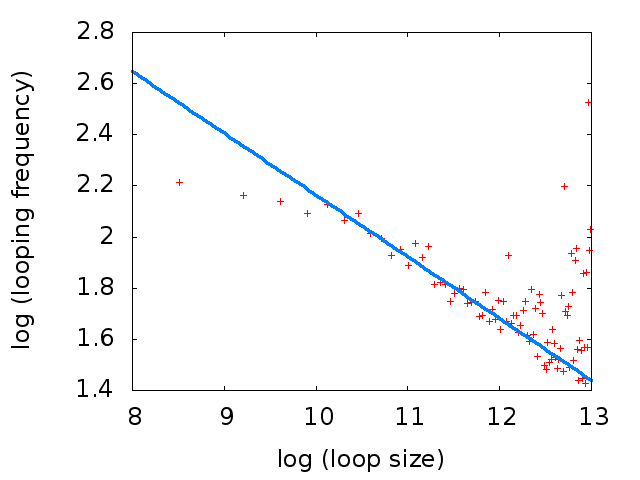}}
\caption{(a) Log-log plot of the ChIA-PET data for CTCF-mediated loops. Solid lines show a fit to a power law for values of the abscissa (natural log of loop size) between $11$ and $13$, and the resulting exponent was $\sim -0.35$. (b) Log-log plot of the ChIA-PET data for contacts associated with RNA Polymerase II; the fit was performed for values of the abscissa (natural log of loop size) between $10$ and $12$, and the resulting exponent was $\sim -0.24$.}
	\label{figChIAPET}
\end{figure}

\section{Captions for Supplementary Movies}

{\bf Suppl. Movie 1:} This movie shows the dynamics corresponding to Suppl. Fig. 4A, where a chromatin fiber of length $L=2000\sigma$ and persistence length $l_p=8\sigma$ is  divided into sections of size $90\sigma$; we assume that each of the sections contains a single slip-link (modelling cohesin) at all times, and that it is delimited by a bead at each boundary which is sticky for the slip-link, to model the presence of CTCF convergent sites. Each of the arcs shown in the movie tracks the positions of the two ends of each slip-link along the chromatin fiber. The interaction between CTCF and cohesin is large enough to ensure virtually irreversible binding on the timescale of our simulations.

{\bf Suppl. Movie 2:} As Suppl. Movie 1, but now with $l_p=4\sigma$, and with sections of size $190\sigma$, with three slip-links per section. It can be seen that the simultaneous presence of the three slip-links leads to a ratcheting effect which favours loop formation.

{\bf Suppl. Movie 3:} This movie shows the self-organization of the osmotic ratchet. The dynamics are slightly different from that shown in Fig.~3 of the main manuscript: now there is not a fixed number of slip-links, but slip-links bind (i.e., are created, when the loading site is unoccupied) at rate $k_{\rm on}=10^{-3}$ s$^{-1}$ and detach (i.e., are destroyed) at rate $k_{\rm off}=10^{-4}$ s${-1}$. The formation of ``rainbow patterns'' with arcs tightly stacked against each other is due to entropic forces which favour the presence of a single loop, kept together by several clustered slip-links, over that of many loops, where slip-links are homogeneously distributed.

\end{document}